\documentclass[10pt]{article}
\usepackage{latexsym,amsmath,amssymb,graphics,amscd}
\usepackage{multirow}
\usepackage{booktabs}
\usepackage{tabularx}
\usepackage[hang,footnotesize]{caption}
\usepackage[pdftex]{graphicx}

\addtocounter{footnote}{1}
\makeatletter
\@addtoreset{figure}{section}
\def\thefigure{\thesection.\@arabic\c@figure}
\def\fps@figure{h,t}
\@addtoreset{table}{bsection}
\def\thetable{\thesection.\@arabic\c@table}
\def\fps@table{h, t}
\@addtoreset{equation}{section}

\makeatother

\newtheorem{theorem}{Theorem}[section]
\newtheorem{definition}[theorem]{Definition}

\newtheorem{remark}[theorem]{Remark}
\newtheorem{proposition}[theorem]{Proposition}

\newcommand{\bfi}{\bfseries\itshape}
\newsavebox{\savepar}

\makeatletter

\begin{document}

\title{\textbf{GARCH options via local risk minimization}}
\author{Juan-Pablo Ortega$^{1}$}
\date{}
\maketitle

\begin{abstract}
We apply a quadratic hedging scheme developed by F\"ollmer, Schweizer, and Sondermann to European contingent products whose underlying asset is modeled using a GARCH process and show that local risk-minimizing strategies with respect to the physical measure do exist, even though an associated minimal martingale measure is only available in the presence of bounded innovations. More importantly, since those local risk-minimizing strategies are in general convoluted and difficult to evaluate, we introduce Girsanov-like risk-neutral measures for the log-prices that yield more tractable and useful results. Regarding this subject, we focus on GARCH time series models with  Gaussian innovations and we provide specific sufficient conditions that have to do with the finiteness of the kurtosis, under which those martingale measures are appropriate in the context of quadratic hedging. When this equivalent martingale measure is adapted to the price representation we are able to recover out of it the classical pricing formulas of Duan~\cite{duan} and Heston-Nandi~\cite{heston nandi}, as well as hedging schemes that improve the performance of those proposed in the literature. 
\end{abstract}

\makeatletter
\addtocounter{footnote}{1} \footnotetext{%
Centre National de la Recherche Scientifique, D\'{e}partement de Math\'{e}%
matiques de Besan\c{c}on, Universit\'{e} de Franche-Comt\'{e}, UFR des
Sciences et Techniques. 16, route de Gray. F-25030 Besan\c{c}on cedex.
France. {\texttt{Juan-Pablo.Ortega@univ-fcomte.fr} }}
\makeatother

\section{Introduction}

GARCH models~\cite{engle arch, bollerslev 86, ding granger engel} have been introduced in the modeling of the time series obtained from financial stock prices with the objective of capturing via a parametric and parsimonious family of processes several features that have been empirically documented and that escape to more elementary modeling tools. For example, the constant variance and drift time series model that one obtains out of the strong Euler discretization of the lognormal model that underlies the Black, Merton, Scholes (BMS) option valuation formula~\cite{black scholes, merton} is not able to account neither for the volatility clustering in the time series of the associated returns nor for the leptokurtosis (fat tails) in their distribution. Moreover, the oversimplification in modeling the stock returns is a source for the appearance of contradictions in the implications of the BMS pricing formula, like the smile shaped curve that one observes when the implied or implicit volatility is plotted as a function of either moneyness or maturity.

From the modeling point of view, the GARCH family is successful at the time of reproducing the above mentioned empirically observed features. Moreover, these models are particularly attractive from the mathematical point of view since the conditions for the existence of stationary solutions can be simply formulated and, additionally, most of the standard techniques in the time series literature concerning model selection and calibration can be adapted to them (see for instance~\cite{gourieroux arch,hamilton ts} and other standard references therein).

The situation becomes more complicated when we try to price contingent products whose underlying asset is assumed to be a realization of a GARCH process. The discrete time character of the model, together with the infinite states space usually assumed on the innovations, makes the associated market automatically incomplete, in the sense that there are payoffs that cannot be replicated using a self-financing portfolio made only out of a bond and the risky asset. This difficulty has been extensively treated in the literature using various approaches.

A first way to address this problem (see~\cite{duan, heston nandi}) consists of adding a term to the GARCH model in the spirit of the NGARCH and VGARCH models introduced in~\cite{engle ng}; when the conditional mean of this modified model is computed using the physical probability, a term proportional to the conditional variance appears that makes it non-risk-neutral. The associated constant of proportionality is interpreted as a return premium per unit of risk. The price of the options that have this model as underlying is then defined as the discounted expected value of the payoff with respect to a new pricing measure under which the GARCH process is risk-neutral; other requirements are also imposed on the pricing measure to ensure, for instance, that the Gaussian character of the innovations is preserved under the new measure. In the case of~\cite{duan}, a utility maximization argument gives legitimacy to this definition.

A different approach consists of finding continuous time processes that extend GARCH to that setup, following a scheme that makes the market complete and yields a price formula in the spirit of the Black-Merton-Scholes theory (see~\cite{kallsen taqqu, kallsen vesenmayer, duan 97} and references therein). It is worth  mentioning that tackling the problem in this way, Kallsen and Taqqu~\cite{kallsen taqqu} obtain results that are consistent with those of~\cite{duan}  as far as the pricing formulas is concerned, but disagree on the associated hedging strategies (see~\cite{garcia renault} for a discussion).

In this paper we will focus on the hedging side of the problem and will implement in the GARCH setting the quadratic hedging approach developed by F\"ollmer, Schweizer, and Sondermann (see~\cite{foellmer sondermann, foellmer schweizer, schweizer 01}, and references therein). Given a probability measure, the theory developed in the papers that we just quoted gives a prescription on the construction of a generalized trading strategy that minimizes the local quadratic hedging risk (to be defined later on). Quadratic hedging techniques can be subjected to improvement since they do not make a difference between hedging shortfalls and windfalls, which should be obviously treated differently as far as the associated risks is concerned. Even though this point has been addressed in a variety of works (see~\cite{pham 2000}, and references therein) the associated hedging and pricing problem is more convoluted;  we will hence put off the use of these techniques in the GARCH context to a future work.

The contents of the paper are organized as follows: Section~\ref{The GARCH family and pricing by  local risk minimization} contains a quick review of the GARCH models  as well as the notions on quadratic hedging that are used later on in the paper.  The last part of this section contains a first result that shows the availability of the quadratic hedging scheme in the GARCH context and a second one where we spell out the conditions under which there exists a {\it  minimal martingale measure}; whenever this measure exists, the value process of the local risk-minimizing strategy (with respect to the {\it physical measure}) admits an interpretation as an arbitrage-free price for the derivative product we are dealing with. Unfortunately, the range of situations in which the minimal martingale measure exists is rather limited and, as we will see, is constrained to GARCH models with bounded innovations; this limitation is, from the modeling point of view, not always appropriate.
Moreover, the expressions that determine the optimal hedging strategy using this measure are convoluted and hence of  limited practical applicability.

The situation that we just described motivates us to carry out in Section~\ref{GARCH with Gaussian and multinomial innovations}  the local risk minimization program for a {\it well chosen} Girsanov-like equivalent martingale measure. We implement this program for GARCH models whose innovations are Gaussian. 
Quadratic hedging with respect to a martingale measure yields much simpler expressions, admits a clear arbitrage-free pricing interpretation and, additionally, the corresponding strategies minimize not only the local risk and the quadratic risk, but also the so-called remaining conditional risk (these concepts will be defined later on). Moreover, we will prove that a linear Taylor expansion in the drift term of the local risk minimizing value process with respect to this martingale measure coincides with the same expansion calculated with respect to the physical measure; consequently, since in most cases the drift term is very small, {\it one can safely compute the risk minimizing strategy with respect to the martingale measure, which is much more convenient, and one obtains practically the same value had one used the much more convoluted expressions in terms of the physical measure}. 

Even though the equivalent martingale measure always exists, the quadratic hedging scheme requires the process modeling the underlying asset to be square summable with respect to the pricing measure. In Theorem~\ref{GARCH martingale measure} we show that a sufficient (but not necessary) condition for that to hold is the finiteness of the kurtosis with respect to the physical measure. It is worth mentioning that with this change of measure, the independent Gaussian innovations of the original GARCH process remain automatically independent and Gaussian after risk neutralization and there is no need to impose this feature as an additional condition (compare with, for example, Assumption 2 in~\cite{heston nandi}). 

The developments that we just explained are formulated using log-prices as the risky asset. In Section~\ref{Local risk minimization in the price representation} we reformulate these results in the price representation using predictible drift terms. This degree of generality allows us to recover in  Section~\ref{Local risk minimization and the pricing formulas of Duan and Heston-Nandi} the classical pricing formulas of Duan~\cite{duan} and Heston-Nandi~\cite{heston nandi} in the context of the local risk minimization scheme. It is worth mentioning that apart from providing an alternative interpretation to existing pricing formulas and improving some theoretical issues (like, for example, dropping assumptions invoked in~\cite{heston nandi} regarding the Gaussian nature of the innovations after risk neutralization), the scheme that we propose comes together with a hedging scheme that performs better than the one proposed in~\cite{duan} and is simply not available in the case of~\cite{heston nandi}. More explicitly, local risk minimizing hedging strategies with respect to a martingale measure minimize by construction the mean square hedging error, when this error is measured with that martingale measure and are hence, from this point of view, preferable to the hedging ratios proposed in~\cite{duan} or~\cite{kallsen taqqu}. Even though there are no theoretical results that guarantee that that this difference in hedging performance still exists when we move to the physical probability, we have carried out numerical tests in Section~\ref{Numerical test of the hedging performance of the local risk minimization scheme} that seem to indicate that this is indeed the case.

\medskip

\noindent {\bf Conventions and notations:} all along this paper we will use the riskless asset as num\'eraire in order not to carry the riskless interest rate in our expressions. Given a filtered probability space $(\Omega, \mathbb{P}, \mathcal{F}, \{ \mathcal{F} _n\}_{n \in \mathbb{N}})$ and $X,Y $ two random variables, we will denote by $E _n[X]:=E[X| \mathcal{F}  _n]$ the conditional expectation, ${\rm cov} _n(X,Y):= {\rm cov}(X,Y| \mathcal{F} _n):=E_n[XY]-E_n[X]E _n[Y]$ the conditional covariance, and by ${\rm var} _n(X):=E_n[X ^2]-E _n[X] ^2 $ the conditional variance. A discrete-time stochastic process $\{X _n\}_{n \in  \mathbb{N}} $ is predictable when $X _n $ is $\mathcal{F} _{n-1} $-measurable, for any $n \in \mathbb{N} $.

\medskip

\noindent\textbf{Acknowledgments:} I thank Josef Teichmann for generously spending his time in going carefully through the paper, giving it a serious thought, and for providing me with relevant feedback that has much improved it. I have also profited from discussions with St\'ephane Chr\'etien, who has shared with me his vast culture in statistics and probability theory. Interesting discussions with Alberto Elices, Friedrich Hubalek, and Andreu L\'azaro are also acknowledged, as well as the comments and criticisms of two anonymous referees that have significantly improved the paper.

\section{The GARCH family and pricing by  local risk minimization}
\label{The GARCH family and pricing by  local risk minimization}

In this section we introduce the general family of times series that we will use for the modeling of the stock prices. We then briefly review the basics of quadratic hedging, and we finally  prove the existence of that kind of strategies in the GARCH context.

\subsection{The GARCH models}

Let $(\Omega, \mathcal{F}, \mathbb{P})$ be a probability space and $\{ \epsilon_n\}_{n \in \mathbb{N}}\sim {\rm IID}(0,1)$ a sequence of zero-mean, square integrable, independent, and identically distributed random variables. We will denote by $\{\mathcal{F}_n\}_{n \in \mathbb{N}} $  the filtration of  $ \mathcal{F}   $ generated by the elements of this family, that is, $\mathcal{F}_n:= \sigma \left( \epsilon _1, \ldots, \epsilon _n\right)$, $n \geq 1 $, is the $\sigma$-algebra generated by $\left\{ \epsilon _1, \ldots, \epsilon _n\right\}$. We will assume that $\mathcal{F}_0 $ is made out of $\Omega   $  and all the negligible events in $\mathcal{F}  $.

GARCH models were introduced in~\cite{bollerslev 86} as a parsimonious generalization of the ARCH models used by Engle~\cite{engle arch} in the modeling of the dynamics of the inflation in the UK. This parametric family has been modified in various forms to make it suitable for the modeling of stock prices and contains an important number of different models. Even though the treatment that we will carry out in the rest of the paper {\it  is valid for all the models in the literature}, we will now pick one of them, namely the one introduced in~\cite{ding granger engel}, to illustrate the features of these models that will be of relevance in the rest of the paper.

\medskip

\noindent {\bf An example: the asymmetric GARCH model:} let $\{S _n\}_{n \in  \mathbb{N}} $ be the sequence that describes the price of the risky asset that we are interested in. The {\bfi  asymmetric GARCH(p,q) model}~\cite{ding granger engel, he terasvirta} determines the dynamics of the prices $\{S _n\}_{n \in  \mathbb{N}} $ by prescribing the following dynamics for the log-returns $r _n:=\log(S _n/S_{n-1})$, that amounts to a recursive relation for the log-prices $s _n:=\log (S _n)$:
\begin{eqnarray}
r _n&=&s _n- s _{n-1}= \mu+ \sigma _n \epsilon _n, \quad \mu\in \mathbb{R},\label{generalized GARCH 1 concrete}\\
\sigma_n ^2&=&\omega+ \sum_{i=1}^p \alpha_i (|\overline{r}_{n-i}|- \gamma  \overline{r}_{n-i})^2+ \sum_{i=1}^q \beta_i \sigma^2_{n-i},\label{generalized GARCH 2 concrete}
\end{eqnarray}
where $\overline{r}_{n}=r_{n}-E[r _n]=r _n- \mu $,  $\{ \epsilon_n\}_{n \in \mathbb{N}}\sim {\rm IIDN}(0,1)$, and $\omega>0 $, $\alpha_i, \beta _i\geq 0 $, $| \gamma |< 1$ are constant real coefficients. Notice that the fact of working with log-prices implies that the price process $\{S _n\}_{n \in  \mathbb{N}} $ determined by~(\ref{generalized GARCH 1 concrete}) and~(\ref{generalized GARCH 2 concrete}) is always positive. The parameter $\gamma$ controls the asymmetric influence of shocks: if they are positive, negative past shocks raise more the variance than comparable positive shocks. This is an empirically observed feature of stock markets.
The following proposition, whose proof is sketched in the appendix, characterizes the constraints on the model parameters that ensure the existence and uniqueness of a weakly stationary solution for~(\ref{generalized GARCH 1 concrete})-(\ref{generalized GARCH 2 concrete}).

\begin{proposition}
\label{stationarity condition statement}
Suppose that $\omega>0 $, $\alpha_i, \beta _i\geq 0 $ and $| \gamma |< 1$. Then the model~(\ref{generalized GARCH 1 concrete})-(\ref{generalized GARCH 2 concrete}) admits a unique weakly (second order) stationary solution if and only if
\begin{equation}
\label{stationarity condition}
(1+ \gamma^2) \left(\alpha_1+ \cdots+ \alpha _p\right)+ \beta_1+ \cdots+ \beta_q< 1,
\end{equation}
in which case
\begin{equation}
\label{variance when stationary}
{\rm var}(r _n)=E[\sigma_n^2]= \frac{\omega}{1-(1+ \gamma^2) \left(\alpha_1+ \cdots+ \alpha _p\right)-\left( \beta_1+ \cdots+ \beta_q\right)}.
\end{equation}
\end{proposition}

\medskip

\noindent {\bf Finite kurtosis:} apart from the second order stationarity studied in the previous statement, it will be important in the developments of the paper to be able to characterize the situations in which the solutions of ~(\ref{generalized GARCH 1 concrete})-(\ref{generalized GARCH 2 concrete}) have finite kurtosis, that is, they have a fourth order moment. This point is relevant for two reasons: first, in the presence of finite kurtosis, it can be shown that the square of a GARCH process is a linear ARMA  model; given that for ARMA there exists a large array of preliminary estimation tools for model selection and calibration, one can take advantage of this situation in the calibration of the GARCH process that one is interested in. Second, as we will see in Theorem~\ref{GARCH martingale measure}, having finite kurtosis is a sufficient condition so that pricing by local risk-minimization is available with respect to  a risk-neutral measure. In the particular case of the asymmetric GARCH model, the existence of finite order moments has been characterized in~\cite{ling mcaleer, li ling mcaleer} in the following compact form: the necessary and sufficient condition for the existence of the moment of order $2m $ is that
\begin{equation}
\label{condition existence moments}
\rho \left[ E \left[ A^{\otimes m}\right]\right] <1,
\end{equation}
where $\otimes $ denotes the Kronecker product of matrices,  $ \rho(B)=\max \left\{ |\text{eigenvalues of the matrix $B$}|\right\} $,  $A$ is the matrix given by
\[
A= \left(
\begin{array}{ccc|ccc}
\alpha_1  Z_t& \cdots & \alpha_p Z_t&\beta_1  Z_t& \cdots & \beta_q Z_t\\
    &I_{(p-1)\times (p-1)}&0_{(p-1)\times 1}& &0_{(p-1)\times q}& \\
\hline
\alpha_1   & \cdots & \alpha_p&\beta_1  & \cdots & \beta_q  \\
    &0_{(q-1)\times p}&  & & I_{(q-1)\times (q-1)}&0_{(q-1)\times 1}
\end{array}
\right),
\]
$I_{p \times p}$ is the $p \times  p $ identity matrix, and $Z _t:= \left(| \epsilon _t|- \gamma\epsilon_t\right) ^2 $. For $m=1 $, the condition~(\ref{condition existence moments}) is the same as~(\ref{stationarity condition}). The kurtosis is finite whenever~(\ref{condition existence moments}) holds with $m=2 $. For example, in the case of a GARCH(1,1) model,~(\ref{condition existence moments}) amounts to the following inequality relation among the model parameters:
\begin{equation*}
\beta^2+ 2 \beta \alpha(1+\gamma ^2)+ 3 \alpha^2\left[(1+ \gamma ^2)^2+ 4 \gamma ^2 \right]<1.
\end{equation*}
The paper~\cite{ling mcaleer 0} contains the corresponding characterization for the finiteness of the kurtosis of other asymmetric GARCH(1,1) processes (like GJR-GARCH) or driven by non-normal innovations.

\medskip

\noindent {\bf Volatility clustering and leptokurtosis:} GARCH is successful in capturing these two features that one empirically observes in stock market log-returns. Actually, in the GARCH context, these two notions are intimately related in the sense that one can say that heteroscedasticity (volatility clustering) causes leptokurtosis (heavy tails) and vice versa. Indeed, since we are using Gaussian innovations, we have
\begin{equation*}
E_{n-1}\left[ \sigma _n^4 \epsilon _n ^4\right]=3 \sigma _n ^4=3 \left(E_{n-1}\left[ \sigma_n^2 \epsilon _n ^2\right] \right)^2.
\end{equation*}
This allows us to write down the kurtosis (standardized fourth moment) as
\begin{eqnarray*}
{\mathcal K}&= & \frac{E\left[ \sigma _n^4 \epsilon _n ^4\right]}{ \left(E\left[ \sigma_n^2 \epsilon _n ^2\right] \right)^2}= \frac{3E\left[ \sigma_n^2 \epsilon _n ^2\right] ^2+3E \left[ \left( E_{n-1}\left[ \sigma_n^2 \epsilon _n ^2\right]\right)^2\right]-3E\left[ \sigma_n^2 \epsilon _n ^2\right] ^2}{\left(E\left[ \sigma_n^2 \epsilon _n ^2\right] \right)^2}\\
    &=&3+3 \frac{E \left[ \left( E_{n-1}\left[ \sigma_n^2 \epsilon _n ^2\right]\right)^2\right]-E \left[ E_{n-1}\left[ \sigma_n^2 \epsilon _n ^2\right]\right]^2}{\left(E\left[ \sigma_n^2 \epsilon _n ^2\right] \right)^2}=3+3 \frac{{\rm var} \left(E_{n-1}\left[ \sigma_n^2 \epsilon _n ^2\right] \right)}{\left(E\left[ \sigma_n^2 \epsilon _n ^2\right] \right)^2}\\
    &= &3+3 \frac{{\rm var} \left(\sigma_n^2\right)}{\left(E\left[ \sigma_n^2 \right] \right)^2},
\end{eqnarray*}
where $E\left[ \sigma_n^2 \right]  $ is determined by~(\ref{variance when stationary}). Notice that this expression, due to~\cite{gourieroux arch}, proves that the excess kurtosis is positive whenever the variance of the volatility is non-zero.

\medskip

\noindent {\bf General GARCH models:} the results that we will prove in this paper apply beyond time series models that follow exactly the functional prescription determined by expressions~(\ref{generalized GARCH 1 concrete}) and~(\ref{generalized GARCH 2 concrete}). In our discussion it will be enough to assume that the log-prices evolve according to:
\begin{eqnarray}
\log \left( \frac{S _n}{S_{n-1}}\right)&=&s _n- s _{n-1}= \mu_n+ \sigma _n \epsilon _n,\label{generalized GARCH 1}\\
\sigma_n ^2&=&\sigma_n ^2(\sigma_{n-1}, \ldots, \sigma_{n-\max (p,q)}, \epsilon _{n-1}, \ldots, \epsilon_{n-q}),
\label{generalized GARCH 2}
\end{eqnarray}
where $\{ \epsilon_n\}_{n \in \mathbb{N}}\sim {\rm IID}(0,\sigma ^2)$, $\{\mu_n\}_{n \in  \mathbb{N}} $ is a predictable process (that is $\mu_n $ is measurable with respect to $\mathcal{F}_{n-1}:= \sigma(\epsilon_1, \ldots, \epsilon_{n-1})$, for all $n \in \mathbb{N} $), and  the  function $\sigma_n ^2(\sigma_{n-1}, \ldots, \sigma_{n-\max (p,q)}, \epsilon _n, \ldots, \epsilon_{n-q}) $  is constructed so that the following two conditions  hold:
\begin{description}
\item [(GARCH1)]  There exists a constant $\omega>0 $ such that $\sigma _n ^2\geq \omega $.
\item [(GARCH2)] The process $\{ \sigma _n\epsilon_n\}_{n \in  \mathbb{N}}$ is weakly (autocovariance) stationary.
\end{description}
A process $\{ s  _n\}_{n \in \mathbb{N}}$ determined by~(\ref{generalized GARCH 2}) and~(\ref{generalized GARCH 2}) will be generically called a GARCH(p,q) process. Notice that~(\ref{generalized GARCH 2}) implies that the time series $\{ \sigma_n\}_{n \in \mathbb{N} }$ is predictable; this feature is the main difference between GARCH and the so-called stochastic volatility models.

\subsection{Local risk minimizing strategies}
\label{Local risk minimizing strategies}

In the following paragraphs we briefly review the necessary concepts on pricing by local risk minimization that we will needed in the sequel. The reader is encouraged to check with Chapter 10 of the excellent monograph~\cite{foellmer schied} for a self-contained and comprehensive presentation of the subject.

Let $H(S _T)$ be a European contingent claim that depends on the terminal value of the risky asset  $S _n $. In the context of an incomplete market, it will be in general impossible to replicate the payoff $H$ by using a self-financing portfolio. Therefore, we introduce the notion of {\bfi  generalized trading strategy}, in which the possibility of additional investment in the num\'eraire asset throughout the trading periods up to expiry time $T$ is allowed. All the following statements are made with respect to a fixed filtered probability space $(\Omega, \mathbb{P}, \mathcal{F}, \{ \mathcal{F} _n\}_{n \in\{0, \ldots, T\}})$.

\begin{definition}
A {\bfi  generalized trading strategy} is a pair of stochastic processes $(\xi^0, \xi)$ such that  $\{\xi^0_n\}_{n \in \{0, \ldots,T\}}$    is adapted and $\{\xi_n\} _{n \in \{1, \ldots,T\}}$ is predictable. The {\bfi  value process} $V $ of $(\xi^0, \xi) $ is defined as
\begin{equation*}
V _0:= \xi_0, \quad \mbox{ and  } \quad V _n:= \xi_n^0+ \xi_n\cdot S _n, \quad n\geq 1.
\end{equation*}
The {\bfi  gains process} $G$ of the generalized trading strategy $(\xi^0, \xi)$ is given by
\begin{equation*}
G _0:=0 \quad \mbox{and } \quad G _n:=\sum_{k=1}^n \xi_k\cdot (S _k-S_{k-1}), \quad n=0, \ldots ,T,
\end{equation*}
and the {\bfi  cost process} $C$ is defined by the difference
\begin{equation*}
C _n:=V _n - G _n, \quad n=0, \ldots, T.
\end{equation*}
\end{definition}
It is easy to check that the strategy $(\xi^0, \xi)$ is self-financing if and only if the value process takes the form
\begin{equation}
\label{self-financing characterization primary}
V _0= \xi_1^0+ \xi_1\cdot  S_0 \quad \mbox{and} \quad V _n =V _0+\sum_{k=1}^n \xi_k \cdot  (S_k-S_{k-1})=V _0+G _n, \quad n=1, \ldots, T,
\end{equation}
or, equivalently, if
\begin{equation}
\label{self-financing characterization}
V _0=C _0= C _1= \ldots =C _T.
\end{equation}

\begin{definition}
\label{admissible setup}
Assume that both $H$  and the $\{S _n\}_{n \in\{0, \ldots, T\}}$ are $L ^2(\Omega, \mathbb{P})$. A generalized trading strategy is called {\bfi  admissible} for $H$ whenever it is in $L ^2(\Omega,\mathbb{P}) $ and its associated value process is such that
\begin{equation*}
V _T=H, \quad {\rm \mathbb{P}}\ {\it a.s.}\quad \mbox{and} \quad V _t \in L ^2(\Omega,\mathbb{P}), \quad \mbox{for each} \ t,
\end{equation*}
and its gain process $G _t \in L ^2(\Omega,\mathbb{P})$, for each $t$.
\end{definition}

The next definition introduces the strategies we are interested in.

\begin{definition}
The {\bfi  local risk process} of an admissible strategy $(\xi^0, \xi)$ is the process
\begin{equation*}
R _t(\xi^0, \xi):=E _t[(C _{t+1}-C _t)^2], \quad t=0, \ldots, T-1.
\end{equation*}
The admissible strategy  $(\widehat{\xi}^0, \widehat{\xi})$ is called {\bfi  local risk-minimizing}  if
\begin{equation*}
R _t(\widehat{\xi}^0, \widehat{\xi})\leq R _t(\xi^0, \xi), \quad {\rm \mathbb{P}}\ {\it a.s.}
\end{equation*}
for all $t$ and each admissible strategy $(\xi^0, \xi)$.
\end{definition}
\bigskip

It can be shown that~\cite[Theorem 10.9]{foellmer schied} an admissible strategy is local risk-minimizing if and only if the cost process is a $\mathbb{P}$-martingale and it is strongly orthogonal to $S $, in the sense that ${\rm cov} _n(S_{n+1}-S _n, C_{n+1}-C _n)=0 $, $\mathbb{P}$-a.s., for any $t=0, \ldots, T-1 $. An admissible strategy whose cost process is a $\mathbb{P}$-martingale is usually referred to as {\bfi  mean self-financing} (recall~(\ref{self-financing characterization}) for the reason behind this terminology). Once a probability measure $\mathbb{P}$ has been fixed, if there exists a local risk-minimizing strategy  $(\widehat{\xi}^0, \widehat{\xi})$ with respect to it, then it is unique (see~\cite[Proposition 10.9]{foellmer schied}) and the payoff $H$ can be decomposed as (see~\cite[Corollary 10.14]{foellmer schied})
\begin{equation}
\label{decomposition payoff risk minimizing}
H=V _0+G _T+L _T,
\end{equation}
with $G _n $ the gains process associated to $(\widehat{\xi}^0, \widehat{\xi})$ and $L _n:=C _n-C _0 $, $n=0, \ldots,T $. Since  $(\widehat{\xi}^0, \widehat{\xi})$ is local risk-minimizing, the sequence $\{L _n\}_{n \in \{0, \ldots, T\}}$, that we will call {\bfi  global (hedging) risk process}, is a square integrable martingale that is strongly orthogonal to $S$ and that satisfies $L _0 =0 $. The decomposition~(\ref{decomposition payoff risk minimizing}) and~(\ref{self-financing characterization primary}) show that $L _T $ measures how far $H $ is from the terminal value of the self-financing portfolio uniquely determined by the initial investment $V _0 $ and the trading strategy $\widehat{\xi} $ (see~\cite[Proposition 1.1.3]{lamberton lapeyre}).

\subsection{Local risk minimization in the GARCH context and minimal martingale measures}

As we pointed out in the previous section, the local risk-minimization aproach to hedging demands picking a particular probability measure in the problem. Given a contingent product on a GARCH driven risky asset, the physical probability measure is the most conspicuous one since, from the econometrics point of view, it is the measure naturally used to calibrate the model.

Our next proposition shows that a local risk-minimizing strategy with respect to the physical measure does exist  in the GARCH context. Given the specific form of~(\ref{generalized GARCH 1}) and~(\ref{generalized GARCH 2}), it is more convenient to reformulate the problem by finding a local risk-minimizing strategy in which we take the log-prices $s _n$ as the risky asset and    $h(s _T):=H(\exp (s _T)) $ as the payoff function.

\begin{proposition}
\label{with respect to physical measure}
Consider a market with a single risky asset that evolves in time according to a GARCH process satisfying~(\ref{generalized GARCH 1}) and~(\ref{generalized GARCH 2}), driven by innovations $\{ \epsilon_n\}_{n \in \mathbb{N}}\sim {\rm IID}(0,\sigma ^2)$. Suppose also that the GARCH process has a bounded drift $\{\mu _n \}_{n \in  \mathbb{N}}$, that is, $\mu_n< B $ for some $B \in \mathbb{R} $ and for all $n \in \mathbb{N} $. Let $h \in L ^2(\Omega, \mathbb{P}, \mathcal{F}_T )$ be a contingent product on $s =\log(S)$. Then, there exists a unique local risk-minimizing strategy for $h$ with respect to the physical measure $\mathbb{P}$, uniquely determined by the following recursive relations
\begin{align}
\widehat{\xi} _k &= \frac{1}{\sigma^2\sigma _k}E_{k-1}\left[h \left(1- \frac{\mu_T}{\sigma^2\sigma_T}\epsilon _T \right) \left(1- \frac{\mu_{T-1}}{\sigma^2\sigma_{T-1}}\epsilon _{T-1} \right)\cdots \left(1- \frac{\mu_{k+1}}{\sigma^2\sigma_{k+1}}\epsilon _{k+1} \right) \epsilon _k\right], \  k=1, \ldots, T-1,\label{local risk minimizing 1}\\
\widehat{\xi} _T &= \frac{1}{\sigma^2\sigma _T}E_{T-1} \left[ h \epsilon _T \right],\label{local risk minimizing 2}\\
V _k &=E_{k}\left[h \left(1- \frac{\mu_T}{\sigma^2\sigma_T}\epsilon _T \right) \left(1- \frac{\mu_{T-1}}{\sigma^2\sigma_{T-1}}\epsilon _{T-1} \right)\cdots \left(1- \frac{\mu_{k+1}}{\sigma^2\sigma_{k+1}}\epsilon _{k+1} \right)\right], \quad k=0, \ldots, T-1,\label{local risk minimizing 3}\\
V _T&= h.\label{local risk minimizing 4}
\end{align}
The position on the riskless asset is given by $\widehat{\xi} _k^0:= V _k- \widehat{\xi}_k  s _k $.
\end{proposition}

\begin{remark}
\normalfont
The condition on the boundedness of the drift is verified for most models considered in the literature in the context of derivatives pricing. In some instances the drift is just a constant (see for example~\cite{barone engle mancini}). Other important situation where this hypothesis trivially holds is the model in~\cite{duan}, where the drift is given by $\mu_n:=r+ \lambda \sigma _n- \frac{1}{2} \sigma _n ^2 $; $r$ and $\lambda  $ are positive constants that account for the continuously compounded one time step risk-free interest rate and the unit risk premium, respectively. In this situation, it is clear that $\mu _n <r+ \frac{1}{2} \lambda ^2  $ for any $n \in \mathbb{N} $.
\end{remark}

\noindent\textbf{Proof of Proposition~\ref{with respect to physical measure}.\ \ } We start by noticing that since $ \sigma_n ^2 $ is $\mathcal{F} _{n-1} $-measurable, the relations~(\ref{generalized GARCH 1}) and~(\ref{generalized GARCH 2}) imply
\begin{align}
E_{n-1}[ \sigma_n\epsilon_n]&=0,\label{increments for GARCH 1}\\
E_{n-1}[s _n-s_{n-1}]&=  \mu_n,\label{increments for GARCH 2}\\
E_{n-1}[(s _n-s_{n-1}) ^2]&=  \mu_n^2+ \sigma^2 \sigma^2_n,\label{increments for GARCH 3}\\
{\rm var}_{n-1}[s _n-s_{n-1}]&=  \sigma^2 \sigma^2_n,\label{increments for GARCH 4}
\end{align}
for any $n \in \{1, \ldots, T\} $.

The first fact that we need to check is that the GARCH  context fits the framework established by Definition~\ref{admissible setup} to carry out hedging by local risk minimization. More explicitly, we have to verify that the log-prices $s$ are square integrable. This is a consequence of hypothesis  {\bf (GARCH2)}; indeed, for any $n \in \{1, \ldots, T\} $,
$
s _n= s _0+ \sum_{i=1}^n\mu_i+ \sigma _1 \epsilon_1+ \cdots + \sigma _n \epsilon _n $. Then,
\begin{equation*}
E[s _n ^2]=E\left[\left(s _0+ \sum_{i=1}^n\mu_i\right)^2+ \sum_{i=1}^n \sigma _i ^2 \epsilon _i ^2+ 2 \sum_{i<j=1}^n
\sigma _i \sigma _j \epsilon _i \epsilon _j\right].
\end{equation*}
Let $i<j $, then $E\left[\sigma _i \sigma _j \epsilon _i \epsilon _j\right]=E[E_{j-1}[\sigma _i \sigma _j \epsilon _i \epsilon _j]]=E[\sigma _i \sigma _j \epsilon _iE_{j-1}[ \epsilon _j]]=E[\sigma _i \sigma _j \epsilon _iE[ \epsilon _j]]=0$. Since by hypothesis  {\bf (GARCH2)} $E[\sigma_i ^2 \epsilon_i ^2]< \infty $ and the drift is bounded, we have that
\begin{equation*}
E[s _n ^2]=E\left[\left(s _0+ \sum_{i=1}^n\mu_i\right)^2  \right]+ \sum_{i=1}^nE[\sigma_i ^2 \epsilon_i ^2]\leq \left(s _0+ nB\right)^2+ \sum_{i=1}^nE[\sigma_i ^2 \epsilon_i ^2]< \infty,
\end{equation*}
as required.

Now, according to~\cite[Proposition 10.10]{foellmer schied}, the existence and uniqueness of a local risk-minimizing strategy is guaranteed as long as we can find a constant $C$ such that $(E_{n-1}[s _n-s_{n-1}]) ^2\leq C \cdot {\rm var}_{n-1}[s _n-s_{n-1}] $, $\mathbb{P}$-a.s. for any $n$. In our case it suffices to take $C= B^2/ ( \sigma^2\omega) $.  Indeed, with this choice and using~(\ref{increments for GARCH 2}) and~(\ref{increments for GARCH 4}),
\begin{equation}
\label{holds with c}
\frac{(E_{n-1}[s _n-s_{n-1}]) ^2}{{\rm var}_{n-1}[s _n-s_{n-1}]}= \frac{ \mu_n^2}{\sigma^2\sigma _n ^2}\leq \frac{B^2}{\sigma^2\omega}=C,
\end{equation}
as required. The recursions~(\ref{local risk minimizing 1})-(\ref{local risk minimizing 4}) follow by rewriting expression~(10.5) in~\cite{foellmer schied} using  the equalities~(\ref{increments for GARCH 1})-(\ref{increments for GARCH 4}).
\quad $\blacksquare$

\medskip

Expressions~(\ref{local risk minimizing 1})-(\ref{local risk minimizing 4}) are convoluted and difficult to evaluate. Moreover, expression~(\ref{local risk minimizing 3}) does not allow us to interpret $V _k $ as an arbitrage free price for $h$ at time $k$.  There are  two possibilities to go around this problem: the first one consists of dropping the physical probability and of choosing instead an equivalent martingale measure that has particularly good properties that make it a legitimate proxy for the original measure. This is the path that we will take in the next section. 

As an alternative, one may want to look for an equivalent martingale measure for which the value process of the local risk-minimizing strategy {\it with respect to the physical measure} can be interpreted as an arbitrage free price for $h$. This is the motivation for introducing the so-called {\bfi  minimal martingale measure}. This measure  is defined as a martingale measure $\widehat{\mathbb{P}} $ that is equivalent to the physical probability $\mathbb{P} $ and satisfies the following two conditions: $E\left[ \left( d\widehat{\mathbb{P}}/ d \mathbb{P}\right)^2\right]< \infty $ and every $\mathbb{P}$-martingale $M \in L ^2(\Omega,\mathbb{P})$ that is strongly orthogonal to the price process $s$, is also a $\widehat{\mathbb{P}} $-martingale. This measure satisfies an entropy minimizing property~\cite[Proposition 3.6]{schweizer 01} and if $\widehat{E} $ denotes the expectation with respect to $\widehat{\mathbb{P}} $, then  the value process $V _k $ in~(\ref{local risk minimizing 3}) can be expressed as (see Theorem 10.22 in~\cite{foellmer schied})
\begin{equation*}
V _k=\widehat{E}_k [h],
\end{equation*}
which obviously yields the interpretation that we are looking for.

As we see in the next proposition, minimal martingale measures exist in the GARCH context only when the innovations are bounded (for example, when the innovations are multinomial) and certain inequalities among the model parameters are respected.

\begin{proposition}
\label{minimal measure martingale}
Using the same setup as in Proposition~\ref{with respect to physical measure}, suppose that  the innovations in the GARCH model are bounded, that is, there exists $K>0 $ such that $ \epsilon_k< K $, for all $k=1, \ldots ,T $, and that this bound is such that $K< \sigma^2 \sqrt{\omega}/ B $, with $\omega>0 $  the constant such that $\sigma _k ^2\geq \omega $ (see condition {\bf (GARCH1)}) and $B\in  \mathbb{R} $ the upper bound for the drift. Then, there exists a unique minimal martingale measure $\widehat{\mathbb{P}} $ with respect to $\mathbb{P}$. Conversely, if there exists a minimal martingale measure then the innovations in the model are necessarily bounded.

Whenever the minimal martingale measure exists, its Radon-Nikodym derivative is given by 
\begin{equation}
\label{randon nikodym of minimal}
\frac{ d\widehat{\mathbb{P}}}{ d \mathbb{P}}=\prod_{k=1}^T \left(1- \frac{\mu_k \epsilon _k}{\sigma ^2 \sigma_k} \right). 
\end{equation} 
\end{proposition}

\noindent\textbf{Proof.\ \ } We start by recalling that in the proof of Proposition~\ref{with respect to physical measure}, we showed in~(\ref{holds with c}) the existence of a constant $C$ such that 
\[ 
(E_{n-1}[s _n-s_{n-1}]) ^2\leq C \cdot  {\rm var}_{n-1}[s _n-s_{n-1}]
\]
for all $t=1, \ldots, T $. In view of this and Theorem 10.30 in~\cite{foellmer schied}, the existence and uniqueness of a minimal martingale measure $\widehat{\mathbb{P}} $ is guaranteed provided that the following inequality holds
\begin{equation}
\label{when minimal martingale measure exists}
(s _n-s_{n-1})E_{n-1}[s _n-s_{n-1}]<E_{n-1}[(s _n-s_{n-1})^2].
\end{equation}
By~(\ref{increments for GARCH 2}) and~(\ref{increments for GARCH 3}), this inequality is equivalent to $\epsilon_n< \sigma^2 \sigma _n/ \mu_n $ and it obviously holds if the innovations are bounded and the bound satisfies $K< \sigma^2 \sqrt{\omega}/ B$. Conversely, suppose that there exists a minimal martingale measure; Corollary 10.29 in~\cite{foellmer schied} implies that~(\ref{when minimal martingale measure exists}) holds and hence so does $\epsilon_n< \sigma^2 \sigma _n/ \mu_n $. Given that $\sigma _n $ and $\mu_n $  are $\mathcal{F}_{n-1} $ measurable and $\epsilon _n$ is $\mathcal{F} _n$ measurable, this equality can only possibly hold whenever the innovations $\epsilon _n $ are bounded. 

As to expression~(\ref{randon nikodym of minimal}), it is a consequence of  Corollary 10.29 and Theorem 10.30 in~\cite{foellmer schied}. According to those two results, the density  $d\widehat{\mathbb{P}}/ d \mathbb{P} $  is the evaluation at $T$ of the $\mathbb{P}$-martingale 
\begin{equation*}
Z _t := \prod_{k=1} ^t \left( 1+ \lambda _k \cdot  \left( y _k- y_{k-1}\right)\right),
\end{equation*}
where $\lambda_k:=-E_{k-1}[s _k-s_{k-1}]/  {\rm var}_{k-1}[s _k-s_{k-1}]$ and $y _k $ is the martingale part in the  Doob decomposition of $s _k $ with respect to $\mathbb{P}$. Therefore, we have 
\begin{equation*}
\label{doob of sn}
y _k-y_{k-1}=s _k-s_{k-1}-E_{k-1}[s _k-s_{k-1}].
\end{equation*}
Using~(\ref{increments for GARCH 2}) and~(\ref{increments for GARCH 4}) in these expressions the result follows. \quad $\blacksquare$

\section{GARCH with Gaussian innovations}
\label{GARCH with Gaussian and multinomial innovations}

The hedging strategies that come out of~(\ref{local risk minimizing 1})-(\ref{local risk minimizing 4}) are, in general, difficult to compute either explicitly or by Monte Carlo methods. Moreover, the interpretation of the values of the resulting local risk-minimizing portfolio as an arbitrage free price for $h$ needs of a minimal martingale measure whose existence is not always available.

The approach that we take in this section consists of dropping the physical measure and of carrying out the local risk minimization program for a {\it well chosen} Girsanov-like equivalent martingale measure; we will justify later on that this measure can be used as a legitimate proxy for the physical probability. We will implement this program for GARCH models whose innovations are Gaussian, for which no minimal martingale measure exists, according to Proposition~\ref{minimal measure martingale}.

The use of a martingale measure for local risk minimization is particularly convenient since the formulas that determine the generalized trading strategy are particularly simple and admit a clear interpretation. Indeed, it is easy to show that, when written with respect to a martingale measure, the local risk-minimizing strategy is determined by
\begin{align}
\widehat{\xi} _k &= \frac{1}{\sigma^2\sigma _k}E_{k-1}\left[h \left(\mu+ \sigma_k \epsilon _k \right)\right], \  k=1, \ldots, T,\label{local risk minimizing martingale1}\\
V _k &=E_{k}\left[h\right], \quad k=0, \ldots, T.\label{local risk minimizing martingale 2}
\end{align}
The position on the riskless asset is given by $\widehat{\xi} _k^0:= V _k- \widehat{\xi}_k  s _k $. Moreover, local risk-minimizing trading strategies computed with respect to a martingale measure also minimize~\cite[Proposition 10.34]{foellmer schied} the so called {\bfi  remaining conditional risk}, defined as the process $R ^R _t (\xi^0, \xi) :=E_t[(C _T-C _t)^2]$, $t=0, \ldots, T $; this is in general not true outside the martingale setup (see~\cite[Proposition 3.1]{schweizer 01} for a counterexample). 

As we will see in Proposition~\ref{argument why martingale}, apart from the computational convenience and the other arguments provided above, the chosen equivalent martingale measure has a particular legitimacy since a linear Taylor expansion in the drift term of the local risk minimizing value process with respect to this measure coincides with the same expansion calculated with respect to the physical measure; consequently, since in most cases the drift term is very small, carrying out the risk minimizing program with respect to the physical measure or the equivalent martingale one that we introduce below yields virtually the same results.

Consider a GARCH process driven by Gaussian innovations, that is, $\{ \epsilon_i\}_{i \in \{1, \ldots, T\}}\sim {\rm IIDN}(0,1)$.
Since our intention is carrying out the quadratic hedging program, a challenge at the time of finding an equivalent martingale measure consists of making sure that, after the change of measure, we do not leave the square-summable category; as we will see in our next theorem this will be ensured by working with processes with finite kurtosis. 

Moreover, it is desirable that the innovations do not lose the Gaussian character in the new picture; this condition is sometimes imposed as a hypothesis (see, for example, Assumption 2 in~\cite{heston nandi}). In the next theorem, this is naturally obtained as a consequence of the construction. The proof of the following result can be found in the appendix.

\begin{theorem}
\label{GARCH martingale measure}
Let $(\Omega,\mathbb{P}, \mathcal{F} )$ be a probability space. Let $\{s _0, s _1, \ldots, s_T\} $ be a GARCH process determined by a recursive relation of the type~(\ref{generalized GARCH 1})-(\ref{generalized GARCH 2}) and where the innovations $\{ \epsilon_i\}_{i \in \{1, \ldots, T\}}\sim {\rm IIDN}(0,1)$; let $ \mathcal{F} _i:= \sigma(\epsilon_1, \ldots, \epsilon_i)$ be the associated filtration of $\mathcal{F}$. Then,
\begin{description}
\item [(i)] The process
\begin{equation*}
Z _n:=\prod_{k=1}^n \exp \left( - \frac{\mu_k}{\sigma _k}\epsilon _k\right)\exp \left( - \frac{1}{2} \frac{\mu_k ^2}{\sigma _k ^2}\right), \quad n=1, \ldots, T,
\end{equation*}
is a $\mathbb{P}$-martingale. If the drift process $\{ \mu_n\}_{n \in  \mathbb{N} }$ is bounded then $\{ Z_n\}_{n \in  \mathbb{N} }$ is square summable.
\item [(ii)] $Z _T $ defines an equivalent measure $Q$ such that  $Z _T= \frac{ dQ}{d\mathbb{P}} $.
\item [(iii)] The process
\begin{equation}
\label{new innovations}
\widetilde{ \epsilon} _n:= \epsilon_n+ \frac{\mu_n}{\sigma _n}, \quad n=1, \ldots , T,
\end{equation}
forms a ${\rm IIDN}(0,1)$ noise with respect to the new probability $Q$.
\item [(iv)] The log-prices  $\{s _0, s _1, \ldots, s_T\} $ form a martingale with respect to $Q$ and they are fully determined by the relations
\begin{eqnarray}
s _n &= &s _0+ \sigma_1 \widetilde{\epsilon} _1+ \cdots +\sigma_n \widetilde{\epsilon} _n,\label{GARCH with Q 1}\\
\sigma_n ^2&=&\widetilde{\sigma}_n ^2(\sigma_{n-1}, \ldots, \sigma_{n-\max (p,q)}, \widetilde{\epsilon} _{n-1}, \ldots,\widetilde{ \epsilon}_{n-q}).\label{GARCH with Q 2}
\end{eqnarray}
The functions $\widetilde{\sigma} _n ^2 $ are the same as $\sigma_n ^2 $ in~(\ref{generalized GARCH 2}) with $\epsilon _{n-1}, \ldots, \epsilon_{n-q} $ written as a function of  $\widetilde{\epsilon} _{n-1}, \ldots,\widetilde{ \epsilon}_{n-q} $ using~(\ref{new innovations}). If the process $\{ \sigma_n\epsilon_n\}_{n \in  \{1, \ldots,T\}}$ is chosen so that it has finite kurtosis with respect to $\mathbb{P}$ and the drift process $\{ \mu_n\}_{n \in  \mathbb{N} }$ is bounded, then the martingale $\{s _0, s _1, \ldots, s_T\} $ is square integrable with respect to $Q$.
\item [(v)] The random variables in the process $\{ \sigma _i \widetilde{\epsilon} _i\} _{i\in \{1, \ldots, T\}}$ are zero mean and uncorrelated with respect to $Q$.
\end{description}
\end{theorem}

\begin{remark}
\normalfont
The conclusion in part {\bf (iv)} on the $Q$-square integrability of  the martingale $\{s _0, s _1, \ldots, s_T\} $ is very important since it spells out clearly sufficient (but not necessary!) conditions, namely, finite kurtosis and bounded drift under which we are entitled to use this new measure theoretical representation of the problem to compute hedging strategies via local risk minimization.
In some market conditions, the finiteness of the kurtosis may be a rather restrictive condition but it is well characterized in the GARCH context  (see~\cite{li ling mcaleer, ling mcaleer 0, ling mcaleer} and references therein).

The condition on the finiteness of the kurtosis can be weakened to requiring the process $\{ \sigma_n\epsilon_n\}_{n \in  \{1, \ldots,T\}}$ to belong to $L^{2+ \epsilon}(\Omega,\mathbb{P}, \mathcal{F})$, with $\epsilon>0 $ arbitrarily small. This result follows from using in the proof (available in the appendix) the fact that the elements of the process $\{ Z_n\}_{n \in  \{1, \ldots,T\}}$ do actually belong to $L^{q}(\Omega,\mathbb{P}, \mathcal{F})$, for any  $q< \infty $ and by replacing the Cauchy-Schwarz inequality in~(\ref{kurtosis and schwarz}) by H\"older's inequality.
\end{remark}

\noindent {\bf The local risk-minimizing strategy associated to the martingale measure.}
Given a European claim $H(S _T)$ on the risky asset $S$, the martingale measure that we described in the previous theorem can be used to come up with a local risk-minimizing strategy by recasting the problem as a hedging problem where we consider the  log-prices $s _n$ as the risky asset and    $h(s _T):=H(\exp (s _T)) $ as the payoff function.

Suppose that the process $\{ \sigma _n\epsilon_n\}_{n \in \{0, \ldots,T\}} $ has finite kurtosis with respect to the physical measure $\mathbb{P}$ and that the drift is bounded. Part {\bf (iv) } of Theorem~\ref{GARCH martingale measure} guarantees in that situation that the log-prices $\{s _0, \ldots, s _T\}$ are square integrable martingales with respect to $Q$ and hence the local risk minimization approach to hedging applies in this transformed setup. A straightforward computation using the elements in Theorem~\ref{GARCH martingale measure}, shows that, for any $n \in \{1, \ldots, T\} $,
\begin{equation*}
\widetilde{E}_{n-1}[s _n]=s_{n-1}, \quad \widetilde{E}_{n-1}[(s _n-s_{n-1}) ^2]=\widetilde{E}_{n-1}[(\sigma _n \widetilde{\epsilon} _n) ^2]= \sigma _n^2, \quad \mbox{and } \quad
 \widetilde{{\rm var}}_{n-1}[s _n-s_{n-1}]= \sigma _n^2.
\end{equation*}
With these elements, the general local risk-minimizing strategy described in~(\ref{local risk minimizing 1})-(\ref{local risk minimizing 4}) becomes, with the use of this measure:
\begin{align}
\widetilde{V} _k &=\widetilde{E}_{k}[h(s _T)]\label{local risk minimizing martingale 1}, \quad k=0, \ldots, T,\\
\widehat{\xi} _k &= \frac{1}{\sigma _k}\widetilde{E}_{k-1}\left[\widetilde{\epsilon} _k\widetilde{V} _k \right]= \frac{1}{\sigma _k}\widetilde{E}_{k-1}\left[\widetilde{\epsilon} _k\widetilde{E}_{k}[h(s _T)] \right]= \frac{1}{\sigma _k}\widetilde{E}_{k-1}\left[\widetilde{\epsilon} _kh(s _T) \right], \quad  k=1, \ldots, T,\label{local risk minimizing martingale 2}\\
L _T&= C _T-C _0=h(s _T)-\widetilde{V} _0-\sum_{k=1}^T \widehat{\xi}_k(s _k- s_{k-1})=
h(s _T)-\widetilde{E}[h(s _T)]-\sum_{k=1}^T\widetilde{ \epsilon} _k\widetilde{E}_{k-1}\left[\widetilde{\epsilon} _kh(s _T) \right].\label{local risk minimizing martingale 3}
\end{align}
The position on the riskless asset is given by $\widehat{\xi} _k^0:= \widetilde{V} _k- \widehat{\xi}_k  s _k $.

\medskip

We conclude this section by showing that, since in practice the trend term $\mu_n$ is usually very small when the time scale is days or weeks\footnote{As an example consider the drift term $\mu $ corresponding to the standard GARCH models with {\it constant} drift term historically calibrated to the daily log-returns of the following indices between the dates January, 2nd 2007--December 31st, 2008: Dow Jones Industrial Average: $-6.99 \cdot 10^{-4} $, Nasdaq Composite: $-8.53 \cdot 10^{-4} $, S\&P 500: $-8.94 \cdot 10^{-4} $, Euronext 100: $-1.1 \cdot 10^{-3} $.}, the value process for the derivative product $h$ obtained by risk minimization using the martingale measure that we just introduced and the one computed using the physical measure, are very close. We make this explicit in the following statement, whose proof is provided in the appendix.
\begin{proposition}
\label{argument why martingale}
Let $h$ be a derivative product whose underlying asset is modeled using a GARCH process with finite kurtosis and constant drift. Let $V _k $ be the value process~(\ref{local risk minimizing 3}) of the local risk minimizing strategy associated to  $h$ computed with the physical probability. Let $\widetilde{V }_k  $ be  the value process~(\ref{local risk minimizing martingale 1}), this time computed with respect to the martingale measure introduced in Theorem~\ref{GARCH martingale measure}. The linear Taylor expansions of $ V _k $  and $\widetilde{V }_k  $ in the drift term $\mu  $ coincide.
\end{proposition}

\subsection{The martingale measure in the price representation and local risk minimization}
\label{Local risk minimization in the price representation}

The martingale measure presented in Theorem~\ref{GARCH martingale measure} is constructed so that, after risk neutralization, log-prices become martingales driven by IIDN innovations. Obviously, this feature does not guarantee that the prices themselves share this attribute. However, Theorem~\ref{GARCH martingale measure} can be easily modified so that an equivalent result is available for prices. Indeed, consider a European derivative product with payoff function $H$, whose underlying asset is modeled using a general GARCH process as in~(\ref{generalized GARCH 1})-(\ref{generalized GARCH 2}), driven by IIDN(0,1) innovations, that is
\begin{eqnarray}
\log \left( \frac{S _n}{S_{n-1}}\right)&=&s _n- s _{n-1}= \mu_n+ \sigma _n \epsilon _n,\label{generalized GARCH 1 new}\\
\sigma_n ^2&=&\sigma_n ^2(\sigma_{n-1}, \ldots, \sigma_{n-\max (p,q)}, \epsilon _{n-1}, \ldots, \epsilon_{n-q}),
\label{generalized GARCH 2 new}
\end{eqnarray}
with $\{ \epsilon_n\}_{n \in \mathbb{N}}\sim {\rm IIDN}(0,1)$ and $\{\mu_n\}_{n \in  \mathbb{N}} $  a predictable process. Rewrite~(\ref{generalized GARCH 1 new}) as 
\begin{equation}
\label{rewritten GARCH}
s _n=s_{n-1}+ \mu _n+ \frac{1}{2} \sigma _n^2- \frac{1}{2} \sigma _n^2+ \sigma_n \epsilon _n=s_{n-1}+ \widetilde{\mu} _n- \frac{1}{2} \sigma _n^2+ \sigma_n \epsilon _n,
\end{equation}
where 
\begin{equation}
\label{definition mu tilde}
\widetilde{\mu} _n:= \mu _n+ \frac{1}{2} \sigma _n^2
\end{equation}
and  consider  
\begin{equation*}
\widetilde{\epsilon} _n := \epsilon _n+\frac{ \widetilde{\mu} _n}{ \sigma _n}, \quad n=1, \ldots, T.
\end{equation*}
The proof of Theorem~\ref{GARCH martingale measure} can be mimicked to show that the measure $Q$ determined by the Radon-Nikodym derivative $Z _T:= dQ/d \mathbb{P} $ given by
\begin{equation}
\label{martingale for prices}
Z _T:=\prod_{k=1}^T \exp \left( - \frac{\widetilde{\mu}_k}{\sigma _k}\epsilon _k\right)\exp \left( - \frac{1}{2} \frac{\widetilde{\mu}_k ^2}{\sigma _k ^2}\right)
\end{equation}
is such that the process $\{ \widetilde{\epsilon }_n\}_{n \in \{1, \ldots,T\}} $ is an IIDN(0,1) noise with respect to it. Moreover, using $\{ \widetilde{\epsilon }_n\}_{n \in \{1, \ldots,T\}} $ we have that~(\ref{generalized GARCH 1 new})-(\ref{generalized GARCH 2 new}) become
\begin{eqnarray}
s _n &= & s _{n-1}- \frac{1}{2}\sigma _n^2+ \sigma _n \widetilde{\epsilon} _n,\label{generalized GARCH 1 new new}\\
\sigma_n ^2&=&\sigma_n ^2(\sigma_{n-1}, \ldots, \sigma_{n-\max (p,q)}, \widetilde{\epsilon }_{n-1}, \ldots, \widetilde{\epsilon}_{n-q}).
\label{generalized GARCH 2 new new}
\end{eqnarray}
The importance of this modification is that the price process $\{S _n\}_{n \in \{0, \ldots,T\}} $ forms a martingale with respect to $Q $. Indeed, using the facts that $S _n /S_{n-1}=\exp \left(- \frac{1}{2} \sigma _n ^2+ \sigma _n \widetilde{\epsilon} _n \right) $, $\{ \widetilde{\epsilon }_n\}_{n \in \{1, \ldots,T\}} $ is an IIDN(0,1) noise with respect to $Q$, and that $\sigma_{n }$ is $\mathcal{F}_{n-1}$-measurable, we obtain that
\begin{equation*}
\widetilde{E}_{n-1}\left[ \frac{S _n}{S_{n-1}}\right]=\widetilde{E}_{n-1}\left[\exp \left(- \frac{1}{2} \sigma _n ^2+ \sigma _n \widetilde{\epsilon} _n \right)\right]=\exp \left(- \frac{1}{2} \sigma _n ^2 \right)\widetilde{E}_{n-1}\left[\exp \left( \sigma _n \widetilde{\epsilon} _n \right)\right]=1,
\end{equation*}
as required. The local risk-minimizing strategy associated to this martingale measure is easy to compute and is given by the expressions:
\begin{align}
\widetilde{V} _k &=\widetilde{E}_{k}[H]\label{local risk minimizing martingale 1 new}, \quad k=0, \ldots, T,\\
\widehat{\xi} _k &= \frac{S_{k-1}}{\Sigma _k^2}\widetilde{E}_{k-1}\left[H\left(\exp \left(- \frac{1}{2} \sigma _k ^2+ \sigma _k \widetilde{\epsilon _k} \right)-1 \right)\right], \quad  k=1, \ldots, T,\label{local risk minimizing martingale 2 new}\\
L _T&=H-\widetilde{V} _0-\sum_{k=1}^T \widehat{\xi}_k(S _k- S_{k-1}),\label{local risk minimizing martingale 3 new}
\end{align}
where $\Sigma_{k}^2:= {\rm var}_{k-1}(S _k-S_{k-1})= S_{k-1}^2 \left({\rm e}^{\sigma_k^2} -1\right)$. The position on the riskless asset is given by $\widehat{\xi} _k^0:= \widetilde{V} _k- \widehat{\xi}_k  S _k $.

\subsection{Local risk minimization and the pricing formulas of Duan and Heston-Nandi}
\label{Local risk minimization and the pricing formulas of Duan and Heston-Nandi}

The martingale measure $Q$ introduced in~(\ref{martingale for prices}) can be used to provide an alternative risk-minimization interpretation to the pricing formula in Duan~\cite{duan}, which was introduced using a utility maximization argument. The same holds for the formula introduced by Heston and Nandi in~\cite{heston nandi}; in this case, our interpretation is even more valuable for, as far as I know, there is no other mathematical argument that supports the use of the formula suggested by the authors, other than the fact that the result can be expressed using an extremely convenient closed form expression. Moreover, after the change of measure, an additional hypothesis is needed in~\cite{heston nandi} (see Assumption 2 in page 590) in order to make sure that the innovations in the model remain IIDN after risk neutralization. We are able to drop that hypothesis since the IIDN character of those innovations is in our case part of the thesis of Theorem~\ref{GARCH martingale measure}.

As an additional bonus, in both cases the pricing via local risk-minimization comes together with an associated hedging strategy that does not exist for the model in~\cite{heston nandi} and, as we mentioned in the introduction, has been the object of various conflicting proposals in the   case of Duan's model (see~\cite{garcia renault} for a discussion). 

By construction, the self-financing hedging strategy associated to the generalized local-risk minimizing strategy that we propose is also a mean-variance optimal strategy when the corresponding mean square hedging error is measured with the martingale measure. Nevertheless, the hedging risk is ``felt" with the physical probability measure. Since there is no theoretical argument that proves that the optimality of the hedging ratios that we propose with respect to others in the literature survives when we change to that measure, we will carry out a numerical study in the next section that seems to indicate that this is indeed the case.

The formula~(\ref{local risk minimizing martingale 2 new}) that provides the hedging ratios needs in general to be evaluated via a Monte Carlo simulation. However, it is worth noticing that it is readily available for any payoff $H$, which yields a competitive advantage with respect to sensitivity methods which usually require the computation of derivatives of the option price and may prove to be an extremely convoluted task for exotic derivatives. The proof of the following Proposition can be found in the appendix.

\begin{proposition}
\label{we get duan and heston nandi}
The formulas proposed by Duan~\cite{duan} and Heston and Nandi~\cite{heston nandi} for the pricing of European options that have underlying risky assets modeled via a NGARCH and an asymmetric GARCH process, respectively, coincide with the prices of the local risk-minimizing strategies constructed using the martingale measures introduced in~(\ref{martingale for prices}), associated to those two processes.
\end{proposition}

\subsection{Numerical test of the hedging performance of the local risk minimization (LRM) scheme}
\label{Numerical test of the hedging performance of the local risk minimization scheme}

We just saw in Proposition~\ref{we get duan and heston nandi} that the local risk-minimizing approach recovers the pricing formulas of Duan~\cite{duan} and Heston and Nandi~\cite{heston nandi} and we mentioned that there is a general result (see for example~\cite[Proposition 10.37]{foellmer schied}) that ensures that the associated self-financing trading strategy~(\ref{local risk minimizing martingale 1 new})-(\ref{local risk minimizing martingale 3 new}) is variance optimal, that is,
\begin{equation}
\label{variance optimal lrm}
\widetilde{E}\left[ \left(H-\widetilde{V} _0-\sum_{k=1}^T \widehat{\xi}_k(S _k- S_{k-1}) \right)^2 \right]\leq \widetilde{E}\left[ \left(H-V _0-\sum_{k=1}^T \xi_k(S _k- S_{k-1}) \right)^2 \right], 
\end{equation}
for any self-financing trading strategy $\xi _1, \ldots, \xi_T  $ with initial value $V _0$. Given that the hedging risk is measured with the physical probability, a question that needs to be adressed is the validity of the inequality~(\ref{variance optimal lrm}) with respect to that measure, that is, replacing in it the $Q$-expectation $\widetilde{E} $ by the $\mathbb{P} $-expectation $E $. An obvious continuity argument suggests that that statement is going to hold true provided that the modified drift term $\{ \widetilde{\mu}_n\}_{n \in \mathbb{N}}$ is ``sufficiently small''; however there is no general result that we can invoke and we will hence carry out a numerical experiment to asses the performance we are interested in.

Since there is no proposal for hedging ratios in \cite{heston nandi} we will limit our comparison to the hedging strategy proposal in Duan's model for the European call option (see~\cite[Corollary 2.4]{duan}) and the standard Black-Scholes model.

More specifically, we will consider an European call option whose underlying asset has as price $\{S _n\}_{n \in \{0, \ldots,T\}}$  a realization of the model spelled out in~(\ref{Duan 1})-(\ref{Duan 2}) with p=q=1, $\alpha _0=0.00001 $, $\alpha_1=0.2 $, $\beta _1=0.7 $, and $\lambda=0.01 $. Additionally, the initial value equals $S _0=100 $.

Our goal is comparing the three mean square hedging errors $E\left[ \left(H-\widetilde{V} _0-\sum_{k=1}^T \widehat{\xi}_k(S _k- S_{k-1}) \right)^2 \right] $, $E\left[ \left(H-V^D _0-\sum_{k=1}^T \xi^D_k(S _k- S_{k-1}) \right)^2 \right] $, and $E\left[ \left(H-V^{BS} _0-\sum_{k=1}^T \xi^{BS}_k(S _k- S_{k-1}) \right)^2 \right] $, where:
\begin{itemize}
\item  $(\widetilde{V} _0, \{\widehat{\xi}_k\}_{k \in \{1, \ldots,T\}}) $ is the self-financing trading strategy associated to the local risk-minimization scheme~(\ref{local risk minimizing martingale 1 new})-(\ref{local risk minimizing martingale 3 new}) for Duan's model with the parameter values that we just specified. 
\item $(V^{BS} _0, \{\widehat{\xi}_k\}_{k \in \{1, \ldots,T\}}) $ is the self-financing trading strategy associated to the Black-Scholes scheme for a lognormal model with constant volatility $\sigma $ set equal to the stationary value of the volatility of Duan's model under the physical probability, namely $\sigma^2= \alpha_0/(1- \alpha _1- \beta _1) $.
\item $(V^{D} _0, \{\xi_k^D\}_{k \in \{1, \ldots,T\}}) $ is the self-financing trading strategy given by Duan's scheme and specified in Corollary 2.4 of~\cite{duan}, namely, 
\begin{equation}
\label{hedging formula duan}
\xi_k^D:= \widetilde{E}_{k-1}\left[  \frac{S _T}{S_{k-1}}\boldsymbol{1}_{X _T\geq K}\right] , 
\end{equation}
where $K $ is the exercise price of the European call option in question.  By Proposition~\ref{we get duan and heston nandi} $\widetilde{V} _0= V^{D} $.
\end{itemize}

The mean square hedging errors are computed by simulating one thousand random price paths, using Duan's model with respect to the physical probability~(\ref{Duan 1})-(\ref{Duan 1}), as the error is computed using the $\mathbb{P}$-expectation. Each of these price paths are hedged using the three different schemes listed above; the Duan and the local risk-minimization approaches require the estimation via Monte Carlo of the hedging ratios~(\ref{hedging formula duan}) and~(\ref{local risk minimizing martingale 2 new}) at each time step. These $Q$-expectations are computed using each time ten thousand different random paths simulated with the risk-neutralized version~(\ref{Duan 1 rn})-(\ref{Duan 2 rn}) of Duan's model. The martingale property is extremely important in this setup consequently, before the expectations are computed these paths are modified using the empirical martigale simulation (EMS) technique introduced in~\cite{duan ems, duan ems 1} which, additionally reduces the variance of the estimation. The variance of the estimation of the hedging error is evaluated by repeating randomly one hundred times the estimation of the mean square error. 

\begin{figure}[htp]
\includegraphics[scale=.5]{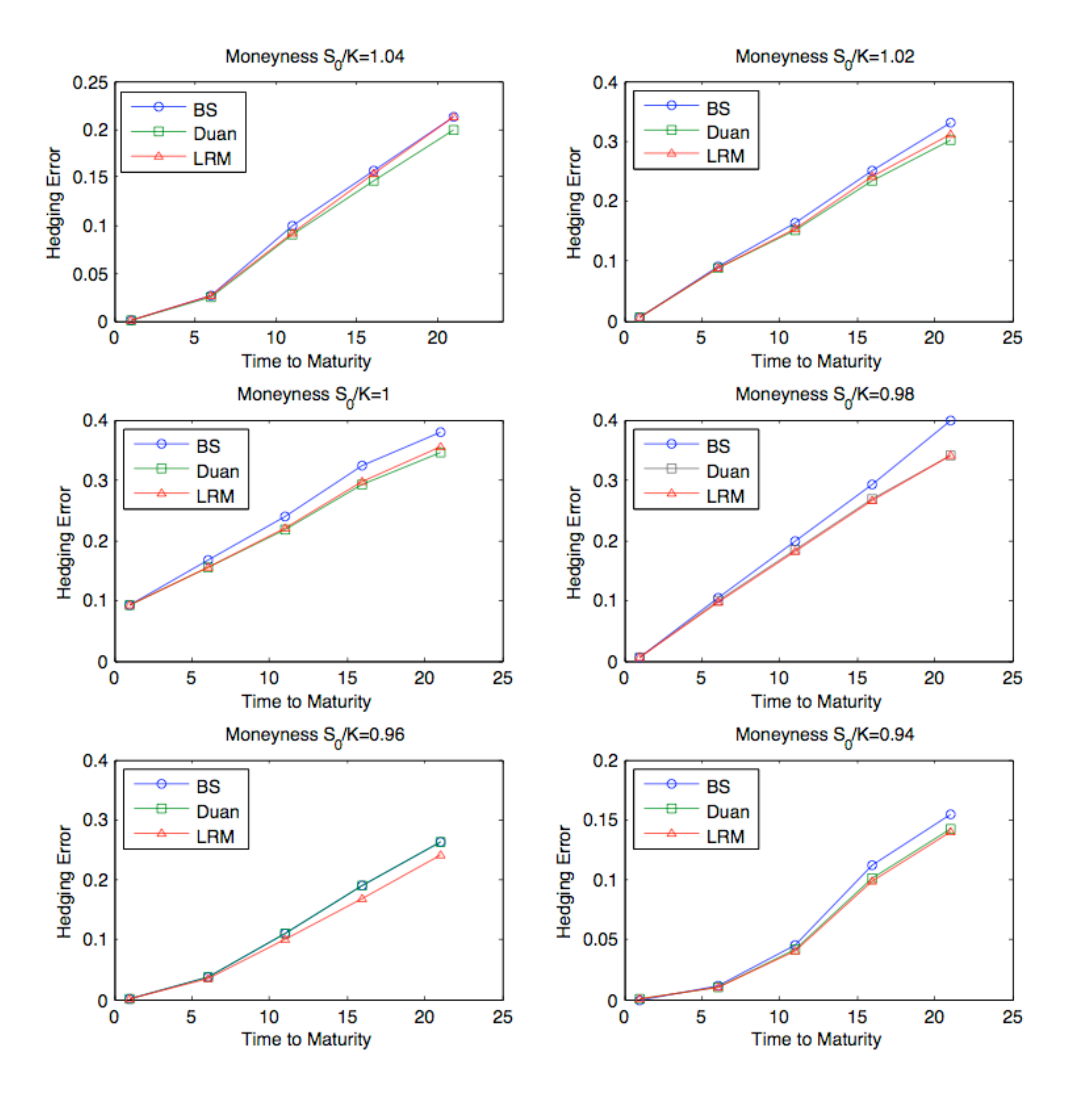}
\caption{Mean square hedging errors for European call options using the Black-Scholes scheme (BS), Duan approach, and local risk-minimzation (LRM).  }
\label{fig:erptsqfit}
\end{figure}

The results of the numerical test are presented in Table~\ref{fig:numerical results} and in Figure~\ref{fig:erptsqfit} and show that:
\begin{itemize}
\item Both Duan's and LRM hedging schemes yield a smaller mean square hedging error than the Black-Scholes scheme.
\item The Duan and LRM hedging schemes produce mean square hedging errors that are not significantly very different, even though Duan performs slightly better than LRM for moneyness bigger than one and LRM performs slightly better than Duan for moneyness smaller than one.
\item The variance of the LRM estimation is always smaller than the one obtained using the Duan and the Black-Scholes schemes. 
\end{itemize}

\begin{table}[!htb]
\vspace{1.5ex}

\noindent\makebox[\textwidth]{%
\begin{tabularx}{1.3\textwidth}{X}
\scalebox{.65}{
\begin{tabular}{ll*{4}{c}*{4}{c}*{4}{c}*{4}{c}}
\toprule
\multicolumn{18}{c}{{\bf Hedging Errors. European Call Options.}} \\
\cmidrule(r){3-18}
	& &\multicolumn{4}{c}{Maturity=6 time steps}&\multicolumn{4}{c}{Maturity=11 time steps}&\multicolumn{4}{c}{Maturity=16 time steps}&\multicolumn{4}{c}{Maturity=21 time steps}\\
\cmidrule(r){3-6}\cmidrule(r){7-10}\cmidrule(r){11-14}\cmidrule(r){15-18}
\multicolumn{2}{l}{$S _0/K $} &1.04 &1&0.98 &0.94&1.04 &1&1.02 &0.94&1.04 &1& 0.98 &0.94&1.04 &1& 0.98 &0.94\\
\midrule
\multirow{2}{*}{{\bf BS}}& \ MSE
	& 0.0262 & 0.1681 &0.1054& 0.0111& 0.0991 &
	0.2402 &0.1978& 0.0457& 0.1573 & 0.3250&0.2924 & 0.1124& 0.2136 & 0.3803 &0.3986& 0.1545\\
	& \ std
	&(0.0082) &(0.0118)&(0.0122) &(0.0071)&(0.0398) &(0.0237)&(0.0214) &(0.0188)&(0.0440) &(0.0271)&(0.0330) &(0.0327)&(0.0344) &(0.0460)&(0.1380) &(0.0285)\\
\midrule
\multirow{2}{*}{{\bf Duan}}& \ MSE
	& 0.0253 & 0.1558&0.0998 & 0.0108& 0.0894 &
	0.2188&0.1844 & 0.0419& 0.1457 & 0.2937&0.2675 & 0.1006& 0.1999 & 0.3465&0.3411 & 0.1427\\
	& \ std
	&(0.0063) &(0.0103)&(0.0106) &(0.0060)&(0.0233) &(0.0243)&(0.0173) &(0.0128)&(0.0290) &(0.0202)&(0.0239) &(0.0204)&(0.0301) &(0.0383)&(0.0428) &(0.0255)\\
\midrule
\multirow{2}{*}{{\bf LRM}}& \ MSE
	& 0.0261 & 0.1544 &0.0979 & 0.0101& 0.0915 &
	0.2201&0.1825 & 0.0409& 0.1536 & 0.2965&0.2673 & 0.0991& 0.2135 & 0.3545&0.3401 & 0.1404\\
	& \ std
	&(0.0060) &(0.0091)&(0.0092) &(0.0052)&(0.0201) &(0.0221)&(0.0162) &(0.0112)&(0.0240) &(0.0184)&(0.0224)&(0.0183) &(0.0275)& (0.0356)&(0.0374) &(0.0240)\\
\midrule
\multicolumn{2}{l}{{\bf std BS vs. Duan}} &30.15\% &14.56\%&15.09\% &18.33 \%&70.81 \% &-2.46 \%&23.69 \% &46.87 \%&51.72 \% &34.15\%& 38.07 \% &60.29 \%&14.28 \% &20.10 \%& 222.42\% &11.76 \%\\	
\midrule
\multicolumn{2}{l}{{\bf std BS vs. LRM}} &36.66\% &29.67\%&32.60\% &36.53 \%&98.00 \% &7.23 \%&32.09 \% &67.87 \%&83.33 \% &47.28\%& 47.32 \% &78.68 \%&25.09 \% &29.21 \%& 268.98 \% &18.75 \%\\	
\midrule
\multicolumn{2}{l}{{\bf std Duan vs.  LRM}} &5.00\% &13.18\%&15.21\% &15.38 \%&15.92 \% &9.95 \%&6.79 \% &14.28 \%&20.83 \% &9.78\%& 6.69 \% &11.47 \%&9.45 \% &7.58 \%& 14.43 \% &6.25 \%\\	
\bottomrule
\end{tabular}}
\end{tabularx}}

\caption{Mean square hedging errors (MSE) for European call options using the Black-Scholes scheme (BS), Duan approach, and local risk-minimization (LRM). The MSEs are computed by simulating $1,000 $ random price paths, using Duan's model with respect to the physical probability; the Duan and LRM hedging ratios for each of those price paths are computed via Monte Carlo using each time $10,000 $ random paths simulated with the risk-neutralized version of Duan's model to which the EMS modification is subsequently applied. The standard deviations (std) are obtained from 100 independent Monte Carlo hedging error estimates. The rows ``std A vs. B'' show the percentage increase in standard deviation experienced in the estimation when using the method A instead of B.}
\label{fig:numerical results}
\end{table}

\section{Conclusions}

In this paper, we studied the applicability of the pricing/hedging scheme by local risk-minimization (LRM) to European options with one underlying asset that is modeled using a GARCH process. The main conclusions of the paper are:
\begin{itemize}
\item Local-risk minimizing strategies exist for options with square summable payoffs even though it is basically only in the presence of models with bounded innovations that a minimal martingale measure is available in order to interpret the subsequent price as an arbitrage-free price.
\item Since the conditions for the existence of a minimal martingale measure are too restrictive, we have carried out the LRM scheme with respect to a well-chosen equivalent martingale measure that produces the same results as the physical probability up to first order in the Taylor series expansion with respect to the drift.
\item When this martingale measure is used in the context of the standard models of Duan~\cite{duan} and Heston and Nandi~\cite{heston nandi} we recover their pricing formulas, which provides an alternative optimal hedging interpretation to the original utility maximization argument that motivated their introduction.
\item The hedging strategy associated to the LRM scheme is additionally mean-variance optimal when the mean square hedging error is computed using the martingale measure.
\item We have studied numerically if this hedging optimality survives when the error is measured using the physical probability by comparing the hedging performances of the standard Black-Scholes scheme, Duan's scheme, and local risk minimization. Simulations show that the Duan and LRM hedging schemes produce mean square hedging errors that are not significantly very different, even though Duan performs slightly better than LRM for moneyness bigger than one and LRM performs slightly better than Duan for moneyness smaller than one. The variance of the LRM estimation is always smaller than the one obtained using both the Duan and the Black-Scholes schemes. 
\item The expression that provides the LRM hedging ratios needs in general to be evaluated via a Monte Carlo simulation but it is readily available for any payoff $H$ and, unlike other hedging proposals based on sensitivity methods, it does not require the computation of derivatives of the option price which may prove to be a very convoluted task for exotic derivatives.
\end{itemize}

\section{Appendix}

\subsection{Proof of Proposition~\ref{stationarity condition statement} }

The proof of the full statement in Proposition~\ref{stationarity condition statement} is lengthy and convoluted. The reader is encouraged to check with~\cite{ling mcaleer, li ling mcaleer}, and references therein. In the following lines we will content ourselves with checking that condition~(\ref{stationarity condition}) implies the asymptotic weak stationarity of the solutions of the model and we will establish~(\ref{variance when stationary}).

We start by noting that $E_{n-1}[r _n]= \mu=E[r _n]$, ${\rm var}_{n-1}(r _n)=E_{n-1}[r _n^2]-E_{n-1}[r _n]^2= \sigma _n^2$, and hence
\begin{equation*}
{\rm var}(r _n)=E \left[ {\rm var}_{n-1} (r _n)\right]+{\rm var} \left(E _{n-1}\left[ r _n\right] \right)=E \left[ \sigma_n^2\right].
\end{equation*}
We now take expectations on both sides of~(\ref{generalized GARCH 2 concrete}), that is,
\begin{equation*}
\sigma_n ^2=\omega+ \sum_{i=1}^p \alpha_i (1+ \gamma^2)\overline{r}_{n-1}^2-2 \gamma \alpha _i|\overline{r}_{n-i}|\overline{r}_{n-i}+ \sum_{i=1}^q \beta_i \sigma^2_{n-i},
\end{equation*}
taking into account that $E[\overline{r}_{n}]=0 $, $E[\overline{r}_{n}^2]=E[\sigma_n^2 \epsilon_n^2]=E[\sigma_n^2]$, and $E[|\overline{r}_{n}|\overline{r}_{n}]=0 $. We obtain
\begin{equation*}
E[\sigma_n^2]= \omega+(1+ \gamma^2)A(L)E[\sigma_n^2]+B(L)E[\sigma_n^2],
\end{equation*}
where $A(L)$ and $B(L)$ are the polynomials $A(z)= \sum_{i=1}^p \alpha _i z ^i $, $B(z)= \sum_{i=1}^q \beta _i z ^i $ on the one-step lag operator $L$. Equivalently,
\begin{equation*}
E[\sigma_n^2]= \omega+\left[ (1+ \gamma^2)A(L)+B(L)\right]E[\sigma_n^2].
\end{equation*}
This difference equation is stable (see, for instance, Proposition 2.2, page 34 in~\cite{hamilton ts}), that is, it admits an asymptotic solution whenever the roots of the polynomial
\begin{equation}
\label{polynomial ar to control}
1-(1+ \gamma ^2)A (z)-B (z)=0,
\end{equation}
lay outside the unit circle, in which case, expression~(\ref{variance when stationary}) clearly holds.
This condition on the roots of~(\ref{polynomial ar to control}) is equivalent to
\begin{equation}
\label{position roots equivalent}
(1+ \gamma ^2)A (1)+B (1)<1,
\end{equation}
which coincides with~(\ref{stationarity condition}). Indeed, if $(1+ \gamma ^2)A (1)+B (1)\geq 1$, we have that since $(1+ \gamma ^2)A (0)+B (0)=0<1 $, then~(\ref{polynomial ar to control}) has necessarily a real root between $0$ and $1$. Conversely, assume that~(\ref{position roots equivalent}) holds and that $z _0$ is a root of~(\ref{polynomial ar to control}) such that $|z _0|<1 $. Then,
\begin{align*}
1&=(1+ \gamma ^2)A (z _0)+B (z _0)=\left |(1+ \gamma ^2)\sum_{i=1}^p \alpha _i z _0 ^i+\sum_{i=1}^q \beta _i z _0 ^i\right |\\
    &\leq (1+ \gamma ^2)\sum_{i=1}^p \alpha _i |z _0| ^i+\sum_{i=1}^q \beta _i |z _0| ^i\leq (1+ \gamma ^2)A (1)+B (1),
\end{align*}
which contradicts our hypothesis. \quad $\blacksquare$

\subsection{Proof of Theorem~\ref{GARCH martingale measure}}

\noindent\textbf{(i)} We start by proving that
\begin{equation}
\label{change is l1}
E[|Z _n|]=E[Z _n]=1, \quad \mbox{for all } \quad n=1, \ldots,T.
\end{equation}
This equality will be needed later on and guarantees that $Z _n \in L ^1(\Omega, \mathbb{P}, \mathcal{F})$. Indeed, let $p (x):= \frac{1}{\sqrt{2 \pi}} \exp (-x ^2/2)$ be the standard normal distribution. Then,
\begin{align*}
E[Z _n] &= \int_{- \infty}^ {+\infty} dx _1 \cdots dx _n\exp \left(-\frac{\mu_1 x _1}{\sigma _1}- \frac{\mu_1^2}{2 \sigma _1^2} \right)p(x _1) \cdots\exp \left(-\frac{\mu_n x _n}{\sigma _n}- \frac{\mu_n^2}{2 \sigma _n^2} \right)p(x _n) \\
    &=  \int_{- \infty}^ {+\infty} dx _1 \cdots dx _{n-1}\exp \left(-\frac{\mu_1 x _1}{\sigma _1}- \frac{\mu_1^2}{2 \sigma _1^2} \right)p(x _1) \cdots\exp \left(-\frac{\mu _{n-1} x _{n-1}}{\sigma _{n-1}}- \frac{\mu _{n-1}^2}{2 \sigma _{n-1}^2} \right)p(x _{n-1})\\
    &\qquad\qquad\int_{- \infty}^{+ \infty}d x _n\exp \left(-\frac{\mu_n x _n}{\sigma _n(x_{n-1}, \ldots, x _1)}- \frac{\mu_n^2}{2 \sigma _n^2(x_{n-1}, \ldots, x _1)} \right)p(x _n).
\end{align*}
Given that
\begin{equation}
\label{integral for girsanov}
\int_{- \infty}^{+ \infty}d x _n\exp \left(-\frac{\mu _nx _n}{\sigma _n(x_{n-1}, \ldots, x _1)}- \frac{\mu_n^2}{2 \sigma _n^2(x_{n-1}, \ldots, x _1)} \right)p(x _n)=1,
\end{equation}
and that we can repeat this integration procedure $n-1 $ times more, we conclude that $E[Z _n]=1 $. We now recall that $\sigma _1, \ldots , \sigma_n,\mu _1, \ldots , \mu_n $, as well as $\epsilon_1, \ldots , \epsilon_{n-1} $ are $\mathcal{F}_{n-1}$ measurable and hence we can write
\begin{align*}
E_{n-1}[Z _n] &= E_{n-1}\left[\prod_{k=1}^n\exp \left( - \frac{\mu_k}{\sigma _k}\epsilon _k\right)\exp \left( - \frac{1}{2} \frac{\mu_k ^2}{\sigma _k ^2}\right) \right]\\
    &=\prod_{k=1}^{n-1}\exp \left( - \frac{\mu_k}{\sigma _k}\epsilon _k\right)\exp \left( - \frac{1}{2} \frac{\mu_k ^2}{\sigma _k ^2}\right)\exp \left( - \frac{1}{2} \frac{\mu_n ^2}{\sigma _n ^2}\right)E_{n-1}\left[\exp \left( - \frac{\mu_n}{\sigma _n}\epsilon _n\right) \right].
\end{align*}
Since $\sigma _n $ and $\mu _n $ are $\mathcal{F}_{n-1}$-measurable and $\epsilon _n $ is independent from $\mathcal{F}_{n-1}$, this can be rewritten as (see, for example, Proposition A.2.5 in~\cite{lamberton lapeyre})
\begin{align*}
E_{n-1}[Z _n] &=Z _{n-1}\exp \left( - \frac{1}{2} \frac{\mu_n ^2}{\sigma _n ^2}\right)\int_{- \infty}^ {+\infty}  \exp \left( - \frac{\mu_n}{\sigma _n}x\right)\, dx=Z_{n-1},
\end{align*}
as required. We conclude by showing that $Z_n $ is square integrable for all $n=1, \ldots,T $ whenever the drift term is bounded. Indeed, let $B \geq 0$ be such that $\mu _n\leq B $, for all $n \in \mathbb{N} $, then 
\begin{align*}
E[Z _n^2] &= \int_{- \infty}^ {+\infty} dx _1 \cdots dx _n\exp \left(-\frac{2\mu_1 x _1}{\sigma _1}- \frac{\mu_1^2}{ \sigma _1^2} \right)p(x _1) \cdots
\exp \left(-\frac{2\mu_n x _n}{\sigma _n}- \frac{\mu_n^2}{\sigma _n^2} \right)p(x _n) \\
    &=  \int_{- \infty}^ {+\infty} dx _1 \cdots dx _{n-1}\exp \left(-\frac{2\mu_1 x _1}{\sigma _1}- \frac{\mu_1^2}{\sigma _1^2} \right)p(x _1) \cdots\exp \left(-\frac{2\mu_{n-1} x _{n-1}}{\sigma _{n-1}}- \frac{\mu_{n-1}^2}{\sigma _{n-1}^2} \right)p(x _{n-1})\\
    &\qquad\qquad\int_{- \infty}^{+ \infty}d x _n\exp \left(-\frac{2\mu_n x _n}{\sigma _n(x_{n-1}, \ldots, x _1)}- \frac{\mu_n^2}{\sigma _n^2(x_{n-1}, \ldots, x _1)} \right)p(x _n).
\end{align*}
Given that
\begin{equation}
\label{inequality thank god}
\int_{- \infty}^{+ \infty}d x _n\exp \left(-\frac{2\mu_n x _n}{\sigma _n(x_{n-1}, \ldots, x _1)}- \frac{\mu_n^2}{\sigma _n^2(x_{n-1}, \ldots, x _1)} \right)p(x _n)=\exp(\mu_n^2/\sigma _n^2)\leq \exp(B^2/ \omega^2) ,
\end{equation}
where the inequality follows from the hypothesis {\bf (GARCH1)} and the bounded character of the drift,  we can conclude that
\[
E[Z _n^2]\leq  \exp(B^2/ \omega^2) \int_{- \infty}^ {+\infty} dx _1 \cdots dx _{n-1}\exp \left(-\frac{2\mu_1 x _1}{\sigma _1}- \frac{\mu_1^2}{\sigma _1^2} \right)p(x _1) \cdots\exp \left(-\frac{2\mu_{n-1} x _{n-1}}{\sigma _{n-1}}- \frac{\mu_{n-1}^2}{\sigma _{n-1}^2} \right)p(x _{n-1}).
\]
Using repeatedly the inequality~(\ref{inequality thank god}) in the previous formula we obtain
\[
E[Z _n^2]\leq  \exp(nB^2/ \omega^2) < +\infty,
\]
as required.

\medskip

\noindent {\bf (ii)} $Z _T $ is by construction non-negative and~(\ref{change is l1}) shows that $E[Z _T]=\mathbb{P}(Z _T>0)=1 $. This guarantees  (see, for example, Remarks after Theorem 4.2.1 in~\cite{lamberton lapeyre}) that $Q$ is a probability measure equivalent to $\mathbb{P}$.

\medskip

\noindent {\bf (iii)} Denote by $ \widetilde{E} $ the expectations with respect to $Q$. Then, for any $u \in \mathbb{R} $ and $n \in \{1, \ldots, T\} $, we will prove that
\begin{equation}
\label{strategy for iidn}
\widetilde{E}_{n-1}[e^{i u \widetilde{\epsilon} _n}]=\widetilde{E}[e^{i u \widetilde{\epsilon} _n}]=e^{-u ^2/2}.
\end{equation}
The first equality in~(\ref{strategy for iidn}) together with Proposition A.2.2 in~\cite{lamberton lapeyre} show that the random variables $\{\widetilde{\epsilon} _1, \ldots, \widetilde{\epsilon} _T \}$ are independent. The second equality, together with the uniqueness theorem for the characteristic function of a random variable (see, for instance, Theorem 4.2 in~\cite{foata fuchs}) shows that the random variables $\{\widetilde{\epsilon} _1, \ldots, \widetilde{\epsilon} _T \}$ are normally distributed under $Q$. Indeed, using the Bayes rule for conditional expectations and part {\bf (i)}, we have
\begin{align*}
\widetilde{E}_{n-1}[e^{i u \widetilde{\epsilon} _n}]&= \frac{1}{E_{n-1}[Z _T]}E_{n-1}[Z _Te^{i u \widetilde{\epsilon} _n}]=\frac{1}{Z _{n-1}}E_{n-1}\left[Z _Te^{i u \left(\epsilon _n+ \frac{ \mu_n}{\sigma _n}\right)}\right]=\frac{Z _{n-1}}{Z _{n-1}}E_{n-1}\left[\frac{Z _T}{Z_{n-1}}e^{i u \left(\epsilon _n+ \frac{ \mu_n}{\sigma _n}\right)}\right]\\
    &= \int_{- \infty}^ {+\infty} dx _n \cdots dx _T\left[\exp \left(-\frac{\mu_n x _n}{\sigma _n}- \frac{\mu_n^2}{2 \sigma _n^2} \right)p(x _n) \cdots\exp \left(-\frac{\mu_T x _T}{\sigma _T}- \frac{\mu_T^2}{2 \sigma _T^2} \right)p(x _T) \right]e^{i u \left(x _n+ \frac{ \mu_n}{\sigma _n}\right)} \\
    &= \int_{- \infty}^ {+\infty} dx _n \exp \left(-\frac{\mu_n x _n}{\sigma _n}- \frac{\mu_n^2}{2 \sigma _n^2} \right)p(x _n)e^{i u \left(x _n+ \frac{ \mu_n}{\sigma _n}\right)}\\
    & \qquad \int_{- \infty}^ {+\infty} dx _{n+1} \exp \left(-\frac{\mu_{n+1} x _{n+1}}{\sigma _{n+1}}- \frac{\mu_{n+1}^2}{2 \sigma _{n+1}^2} \right)p(x _{n+1})\cdots   \int_{- \infty}^ {+\infty}dx _T \exp \left(-\frac{\mu_T x _T}{\sigma _T}- \frac{\mu_T^2}{2 \sigma _T^2} \right)p(x _T) .
\end{align*}
Given that all the integrals
\begin{equation}
\label{integrals equal one}
\int_{- \infty}^ {+\infty}dx _i \exp \left(-\frac{\mu_i x _i}{\sigma _i}- \frac{\mu_i^2}{2 \sigma _i^2} \right)p(x _i)=1,
\end{equation}
the previous expression reduces to
\begin{equation}
\label{integral with fourier transform}
\widetilde{E}_{n-1}[e^{i u \widetilde{\epsilon} _n}]= \int_{- \infty}^ {+\infty} dx _n \exp \left(-\frac{\mu_n x _n}{\sigma _n}- \frac{\mu_n^2}{2 \sigma _n^2} \right)p(x _n)e^{i u \left(x _n+ \frac{ \mu_n}{\sigma _n}\right)}=e^{-u ^2/2}.
\end{equation}
Regarding the second equality in~(\ref{strategy for iidn}), we compute
\begin{align*}
\widetilde{E}[e^{i u \widetilde{\epsilon} _n}]&= \frac{1}{E[Z _T]}E[Z _Te^{i u \widetilde{\epsilon} _n}]=E\left[Z _Te^{i u \left(\epsilon _n+ \frac{ \mu_n}{\sigma _n}\right)}\right]\\
    &= \int_{- \infty}^ {+\infty} dx _1 \cdots dx _T\left[\exp \left(-\frac{\mu_1 x _1}{\sigma _1}- \frac{\mu_1^2}{2 \sigma _1^2} \right)p(x _1) \cdots\exp \left(-\frac{\mu_T x _T}{\sigma _T}- \frac{\mu_T^2}{2 \sigma _T^2} \right)p(x _T) \right]e^{i u \left(x _n+ \frac{ \mu_n}{\sigma _n}\right)} \\
    &=  \int_{- \infty}^ {+\infty} dx _1 \exp \left(-\frac{\mu_1 x _1}{\sigma _1}- \frac{\mu_1^2}{2 \sigma _1^2} \right)p(x _1) \cdots \int_{- \infty}^ {+\infty} dx _n \exp \left(-\frac{\mu_n x _n}{\sigma _n}- \frac{\mu_n^2}{2 \sigma _n^2} \right)p(x _n)e^{i u \left(x _n+ \frac{ \mu_n}{\sigma _n}\right)}\\
    & \qquad \int_{- \infty}^ {+\infty} dx _{n+1} \exp \left(-\frac{\mu_{n+1} x _{n+1}}{\sigma _{n+1}}- \frac{\mu_{n+1}^2}{2 \sigma _{n+1}^2} \right)p(x _{n+1})\cdots   \int_{- \infty}^ {+\infty}dx _T \exp \left(-\frac{\mu_T x _T}{\sigma _T}- \frac{\mu_T^2}{2 \sigma _T^2} \right)p(x _T) .
\end{align*}
Using again~(\ref{integrals equal one}) and the second equality in~(\ref{integral with fourier transform}) we easily obtain that
\begin{equation*}
\widetilde{E}[e^{i u \widetilde{\epsilon} _n}]=e^{-u ^2/2},
\end{equation*}
as required.

\medskip

\noindent {\bf (iv)} Expressions~(\ref{GARCH with Q 1}) and~(\ref{GARCH with Q 2}) follow from substituting~(\ref{new innovations}) in~(\ref{generalized GARCH 1}) and~(\ref{generalized GARCH 2}). Recall now that
\begin{equation*}
E[\sigma _i ^2 \epsilon _i ^2]=E[E_{i-1}[\sigma _i ^2 \epsilon _i ^2]]=E[\sigma _i ^2 E [\epsilon _i ^2]]=E[\sigma _i ^2].
\end{equation*}
Hence, by hypothesis {\bf (GARCH2)}, we have that
\begin{equation}
\label{finite sigma 22}
E[\sigma _i ^2]=E[\sigma _i ^2 \epsilon _i ^2]< \infty.
\end{equation}
Using now~(\ref{finite sigma 22}), part {\bf (i)}, and Bayes law of conditional probability, we have that
\begin{equation}
\label{inequality L2 Z}
\widetilde{E}[|\sigma_i|]= \widetilde{E}[\sigma _i]=E[Z _T \sigma _i]\leq  \left[E[Z _T ^2] \right]^{\frac{1}{2}} \left[E[\sigma _i^2] \right]^{\frac{1}{2}}< \infty.
\end{equation}
Additionally, since by part {\bf (iii)} the innovations $ \widetilde{\epsilon} _i $ are Gaussian with respect to $Q$, we have
\begin{equation*}
\widetilde{E}\left[ |\sigma_i \widetilde{\epsilon} _i|  \right]=\widetilde{E}\left[ \sigma_i \widetilde{E}_{i-1}\left[ | \widetilde{\epsilon} _i |  \right]\right]=\widetilde{E}\left[ \sigma_i \widetilde{E}\left[ | \widetilde{\epsilon} _i |  \right]\right]=\sqrt{\frac{2}{\pi}} \widetilde{E}[|\sigma _i|],
\end{equation*}
which together with~(\ref{inequality L2 Z}) implies that $\widetilde{E}\left[ |\sigma_i \widetilde{\epsilon} _i|  \right] < \infty$. This inequality and~(\ref{GARCH with Q 1}) show that $s _n \in L ^1(\Omega ,Q, \mathcal{F})$. Indeed,
\begin{equation*}
E[|s _n|]=E[|s _0+ \sigma_1 \widetilde{\epsilon} _1+ \cdots +\sigma_n \widetilde{\epsilon} _n|]\leq
E[|s _0|]+E[| \sigma_1 \widetilde{\epsilon} _1|]+ \cdots +E[|\sigma_n \widetilde{\epsilon} _n|]< \infty.
\end{equation*}
Finally,
\begin{equation*}
\widetilde{E}_{n-1}[s _n]= \widetilde{E}_{n-1}[s_{n-1}+ \sigma_n \widetilde{\epsilon} _n]=s_{n-1}+ \sigma _n \widetilde{E}_{n-1}[\widetilde{\epsilon} _n]=s_{n-1},
\end{equation*}
which proves that $\{s _0, s _1, \ldots, s_T\} $ forms a martingale with respect to $Q$. Notice that in the last two equalities of the previous expression we used the conclusion of point {\bf (iii)}.

Suppose now that the variables $\{ \sigma_i\epsilon_i\}_{i \in  \{1, \ldots,T\}}$ have finite kurtosis with respect to $\mathbb{P}$ and that the drift process $\{ \mu_n\}_{n \in  \mathbb{N} }$ is bounded. Then, for each $i\in  \{1, \ldots,T\}$
\begin{equation}
\label{has finite kurtosis}
E \left[ \sigma_i^4\epsilon_i^4\right] < \infty.
\end{equation}
Then, since $E \left[ \sigma_i^4\epsilon_i^4\right] =E \left[ \sigma_i^4E _{i-1}\left[\epsilon_i^4 \right]\right] =3E \left[ \sigma_i^4\right]$, we have that
\begin{equation}
\label{finite sigma 4}
E \left[ \sigma_i^4\right]< \infty.
\end{equation}
We will proceed by showing first that~(\ref{has finite kurtosis}) and~(\ref{finite sigma 4}) imply that
\begin{equation}
\label{finite s 4}
E \left[ s_n^4\right]< \infty
\end{equation}
or, equivalently,
\begin{equation}
\label{finite expanded s n 4}
E \left[ s_n^4\right]=E \left[ \left( \left(s _0+ \sum_{j=1}^n \mu_j\right)^2+ 2 \sum_{i=1}^n\left(s _0+ \sum_{j=1}^n \mu_j\right) \sigma_i \epsilon_i+ \sum_{i=1}^n  \sigma_i ^2 \epsilon_i^2+ 2 \sum _{i<j=1}^n \sigma_i \sigma _j \epsilon _i \epsilon _j
\right)^2\right]< \infty.
\end{equation}
When the square inside the expectation is expanded, some algebra shows that $E \left[ s_n^4\right] $ is a finite sum of real numbers plus terms that, up to multiplication by finite constants, have the form:
\begin{itemize}
\item $E \left[ \sigma _i \epsilon_i\sigma_j \epsilon _j\right] =E \left[ E_{i-1} \left[\sigma _i \epsilon_i\sigma_j \epsilon _j\right]\right]=E \left[ \sigma _i \epsilon_j\sigma_j E \left[\epsilon _i\right]\right] =0$, where we assume, without loss of generality, that $j<i $.
\item $E \left[ \sigma_i ^2\epsilon _i ^2\right]< \infty $, by hypothesis {\bf (GARCH2)}.
\item  $E \left[ \sigma_i^4\epsilon_i^4\right] < \infty$, by~(\ref{has finite kurtosis}).
\item Also by~(\ref{has finite kurtosis}), the terms of the form
\begin{equation}
\label{two squares finite}
E \left[ \sigma _i ^2\epsilon_i^2\sigma_j ^2\epsilon _j^2\right] \leq \left( E \left[ \sigma_i^4\epsilon_i^4\right]\right)^{1/2} \left( E \left[ \sigma_j^4\epsilon_j^4\right]\right)^{1/2}< \infty.
\end{equation}
\item $E \left[ \sigma_i \sigma_j\sigma_k\sigma_l \epsilon_i\epsilon_j \epsilon_k \epsilon _l\right] $. This term is also finite because by~(\ref{two squares finite})
\begin{equation*}
|E \left[ \sigma_i \sigma_j\sigma_k\sigma_l \epsilon_i\epsilon_j \epsilon_k \epsilon _l\right] |\leq  \left(E \left[ \sigma _i ^2\epsilon_i^2\sigma_j ^2\epsilon _j^2\right] \right)^{1/2}\left(E \left[ \sigma _k ^2\epsilon_k^2\sigma_l ^2\epsilon _l^2\right] \right)^{1/2}< \infty.
\end{equation*}
\item Analogous arguments can be used to prove the finiteness of the remaining terms that have the form $E \left[ \sigma _i^3 \epsilon_i^3\sigma_j \epsilon _j\right] $, $E \left[ \sigma _i^2 \epsilon_i^2\sigma_j \epsilon _j \sigma_k \epsilon _k\right] $, $E \left[ \sigma _i^3 \epsilon_i^3\right] $, $E \left[ \sigma _i^2 \epsilon_i^2\sigma_j \epsilon _j\right] $, and  $E \left[ \sigma_i \sigma_j\sigma_k\epsilon_i\epsilon_j \epsilon_k\right] $.
\end{itemize}
This argument establishes~(\ref{finite expanded s n 4}). We now use this relation to conclude that  $\{s _0, s _1, \ldots, s_T\} $ is square integrable with respect to $Q$. Indeed, by part {\bf (i)} of the theorem, $\{Z _n\}_{n \in \{1, \ldots T\}}  $ is a square integrable  martingale and hence
\begin{equation}
\label{kurtosis and schwarz}
\widetilde{E}\left[ s _n ^2\right]= E \left[ Z _T s _n ^2\right]\leq \left( E \left[ Z _T^2\right] \right) ^{1/2}\left( E \left[ s _n^4\right] \right) ^{1/2}< \infty,
\end{equation}
as required.

\medskip

\noindent {\bf (v)}Let $n \in \{1, \ldots, T\} $. Then, by part {\bf (iii)}
\begin{equation*}
\widetilde{E}_{n-1}[\sigma _n \widetilde{\epsilon} _n]=\sigma _n\widetilde{E}_{n-1}[ \widetilde{\epsilon} _n]=\sigma _n\widetilde{E}[ \widetilde{\epsilon} _n]=0.
\end{equation*}
Now, as $\widetilde{E}[\sigma _n \widetilde{\epsilon} _n]= \widetilde{E}[\widetilde{E}_{n-1}[\sigma _n \widetilde{\epsilon} _n]]=0 $, the first statement follows.

Let $j \in \{1, \ldots, T\} $ and assume, without loss of generality, that $j<n $. Then,
\begin{equation*}
\widetilde{E}_{n-1}[\sigma _n \widetilde{\epsilon} _n \sigma _j \widetilde{\epsilon} _j]=
\sigma _n  \sigma _j \widetilde{\epsilon} _j \widetilde{E}_{n-1}[\widetilde{\epsilon} _n]=
\sigma _n  \sigma _j \widetilde{\epsilon} _j \widetilde{E}[\widetilde{\epsilon} _n]=0.
\end{equation*}
Consequently,
\begin{equation*}
{\rm cov}(\sigma _n \widetilde{\epsilon} _n, \sigma _j \widetilde{\epsilon} _j)=\widetilde{E}[\sigma _n \widetilde{\epsilon} _n \sigma _j \widetilde{\epsilon} _j]= \widetilde{E} \left[\widetilde{E}_{n-1}[\sigma _n \widetilde{\epsilon} _n \sigma _j \widetilde{\epsilon} _j]
\right]=0. \quad  \blacksquare
\end{equation*}

\subsection{Proof of Proposition~\ref{argument why martingale}}

Let $ f _k(\mu) $ be the function defined by the value process~(\ref{local risk minimizing 3}) with respect to the physical measure, that is,
\begin{equation*}
f _k(\mu):=E_{k}\left[h \left(1- \frac{\mu}{ \sigma_T}\epsilon _T \right) \left(1- \frac{\mu}{ \sigma_{T-1}}\epsilon _{T-1} \right)\cdots \left(1- \frac{\mu}{ \sigma_{k+1}}\epsilon _{k+1} \right)\right].
\end{equation*}
A straightforward computation shows that 
\begin{equation}
\label{expansion f k}
f_k(0)=E _k[h] \quad \text{and} \quad f ' _k (0)=-\sum_{j=k+1}^T E _k \left[  h \frac{\epsilon _j }{\sigma _j}\right]. 
\end{equation}
Consequently, the linear Taylor approximation $V _k^{lin} $ of $V _k  $ is given by
\begin{equation}
\label{linear approx}
V _k^{lin}=E _k[h] - \mu\sum_{j=k+1}^T E _k \left[  h \frac{\epsilon _j }{\sigma _j}\right].
\end{equation}
Let now $\widetilde{f} _k(\mu) $ be the value process with respect to the martingale measure $Q $ in Theorem~\ref{GARCH martingale measure}. Using the martingale property of the process $Z _n $ that gives us the Radon-Nikodym derivative $ dQ/ d \mathbb{P} $ we have
\begin{eqnarray}
\widetilde{f} _k(\mu)&:= &\widetilde{E} _k[h]= \frac{1}{E _k[Z _T]} E _k[Z _T h]= \frac{1}{Z _k]} E _k[Z _T h]=E _k\left[\frac{Z _T}{Z _k} h\right]\\
	&= &E _k\left[ h\exp \left(- \frac{\mu}{\sigma_T}\epsilon _T- \frac{\mu ^2}{2 \sigma _T ^2}  \right) \cdots  \exp \left(- \frac{\mu}{\sigma_{k+1}}\epsilon _{k+1}- \frac{\mu ^2}{2 \sigma _{k+1} ^2}  \right)\right].
\end{eqnarray} 
A straightforward computation shows that $\widetilde{f} _k(0)= f _k (0)$ and  $\widetilde{f} '_k(0)= f' _k (0)$. Consequently,   $V _k^{lin}= \widetilde{V} _k^{lin} $, as required. \quad $\blacksquare$

\subsection{Proof of Proposition~\ref{we get duan and heston nandi}}

We start with Duan's model which is given by

\begin{eqnarray}
\log \left( \frac{S _n}{S_{n-1}}\right)&=&s _n- s _{n-1}= \lambda \sigma _n- \frac{1}{2} \sigma_n^2+ \sigma _n \epsilon _n,\label{Duan 1}\\
\sigma_n ^2&=&\alpha_0+\sum_{i=1}^q \alpha_i \sigma_{n-i}^2\epsilon_{n-i}^2+\sum_{j=1}^p \beta_j \sigma_{n-j}^2, \quad \{ \epsilon _n\}_{n \in  \mathbb{N}}\sim {\rm IIDN}(0,1)
\label{Duan 2}
\end{eqnarray}
where $\alpha_0>0 $, $ \alpha_i, \beta _j\geq 0 $  and $\sum_{i=1}^q \alpha_i +\sum_{j=1}^p \beta_j<1 $ so that second order stationarity is ensured and the coefficient $\lambda $ is interpreted as a unit risk premium. Using the notation introduced in~(\ref{generalized GARCH 1 new}) and~(\ref{definition mu tilde}) this is a general GARCH model with $\mu_n =\lambda \sigma _n- \frac{1}{2} \sigma_n^2 $  and $\widetilde{\mu}_n =\lambda \sigma _n$. Consequently, by ~(\ref{generalized GARCH 1 new new})-(\ref{generalized GARCH 2 new new}), after risk neutralization this process is given by 
\begin{eqnarray}
\log \left( \frac{S _n}{S_{n-1}}\right)&=&s _n- s _{n-1}=- \frac{1}{2} \sigma_n^2+ \sigma _n \widetilde{\epsilon} _n, \quad \{ \widetilde{\epsilon} _n\}_{n \in  \mathbb{N}}\sim {\rm IIDN}(0,1)\label{Duan 1 rn}\\
\sigma_n ^2&=&\alpha_0+\sum_{i=1}^q \alpha_i \sigma_{n-i}^2(\widetilde{\epsilon}_{n-i}- \lambda)^2+\sum_{j=1}^p \beta_j \sigma_{n-j}^2,
\label{Duan 2 rn}
\end{eqnarray}
and hence, according to~(\ref{local risk minimizing martingale 1 new}), its local risk-minimizing price is given by $V _0= \widetilde{E}[H]$, which coincides with the formula proposed by Duan since the process~(\ref{Duan 1 rn})-(\ref{Duan 2 rn}) is identical to the one obtained in Theorem 2.2 of~\cite{duan} out of his {\it locally risk-neutral valuation relationship}.

As to Heston and Nandi, their process is given by
\begin{eqnarray}
\log \left( \frac{S _n}{S_{n-1}}\right)&=&s _n- s _{n-1}= \lambda \sigma _n^2+ \sigma _n \epsilon _n,\quad \{ \epsilon _n\}_{n \in  \mathbb{N}}\sim {\rm IIDN}(0,1)\label{HN 1},\\
\sigma_n ^2&=&\alpha_0+\sum_{i=1}^q \alpha_i (\epsilon_{n-i}- \gamma _i \sigma _{n-i})^2+\sum_{j=1}^p \beta_j \sigma_{n-j}^2, 
\label{HN 2}
\end{eqnarray}
where $\alpha_0>0 $, $ \alpha_i, \beta _j\geq 0 $  and the roots of the polynomial $x ^p-\sum_{j=1}^p( \beta _i+ \alpha _i \gamma _i^2)x^{p-i} $ lie inside the unit circle
so that second order stationarity is ensured. This time we have that $\mu_n =\lambda \sigma _n^2$,  $\widetilde{\mu}_n =(\lambda+ \frac{1}{2}) \sigma _n ^2$, and 
\begin{equation}
\label{rn innovations hn}
\widetilde{\epsilon} _n:= \epsilon_n+ \left(\lambda + \frac{1}{2} \right) \sigma_n.
\end{equation}
Hence, the risk neutralized version of~(\ref{HN 1})-(\ref{HN 2}) is
\begin{eqnarray}
\log \left( \frac{S _n}{S_{n-1}}\right)&=&s _n- s _{n-1}=- \frac{1}{2} \sigma_n^2+ \sigma _n \widetilde{\epsilon} _n, \quad \{ \widetilde{\epsilon} _n\}_{n \in  \mathbb{N}}\sim {\rm IIDN}(0,1)\label{hn 1 rn}\\
\sigma_n ^2&=&\alpha_0+\sum_{i=1}^q \alpha_i \left(\widetilde{\epsilon}_{n-i}-\left(\lambda+ \gamma _i+ \frac{1}{2} \right)\sigma_{n-i}\right)^2+\sum_{j=1}^p \beta_j \sigma_{n-j}^2,
\label{hn 2 rn}
\end{eqnarray}
which coincides with the one proposed by Proposition 1 of~\cite{heston nandi}. We emphasize that by Theorem~\ref{GARCH martingale measure}, the risk neutralized innovations $\{ \widetilde{\epsilon} _n\}_{n \in  \mathbb{N}}$ are automatically IIDN(0,1) and, unlike in the treatment carried out in~\cite{heston nandi}, no additional assumption is needed. \quad $\blacksquare$

\end{document}